\documentclass[prb,superscriptaddress,floatfix,twocolumn,showpacs]{revtex4-2}
\usepackage{verbatim}
\usepackage{amsmath}
\usepackage{amsfonts}
\usepackage{amssymb}
\usepackage{graphicx}
\usepackage{bm}
\usepackage{color}
\usepackage{hyperref}
\usepackage[active]{srcltx}
\usepackage{cases}
\usepackage{ulem}
\usepackage{multirow}
\usepackage{braket}

\begin{document}
\title{Band structures under non-Hermitian periodic potentials:\\ Connecting nearly-free and bi-orthogonal tight-binding models}
\author{Ken Mochizuki}
\affiliation{Advanced Institute for Materials Research (WPI-AIMR), Tohoku University, Sendai 980-8577, Japan}
\affiliation{Nonequilibrium Quantum Statistical Mechanics RIKEN Hakubi Research Team, RIKEN Cluster for Pioneering Research (CPR), 2-1 Hirosawa, Wako 351-0198, Japan}

\author{Tomoki Ozawa}
\affiliation{Advanced Institute for Materials Research (WPI-AIMR), Tohoku University, Sendai 980-8577, Japan}

\begin{abstract}
We explore band structures of one-dimensional open systems described by periodic non-Hermitian operators, based on continuum models and tight-binding models. 
We show that imaginary scalar potentials do not open band gaps but instead lead to the formation of exceptional points as long as the strength of the potential does not exceed a threshold value, which is contrast to closed systems where real potentials open a gap with infinitesimally small strength. 
The imaginary vector potentials hinder the separation of low energy bands because of the lifting of degeneracy in the free system.  
In addition, we construct tight-binding models through bi-orthogonal Wannier functions based on Bloch wavefunctions of the non-Hermitian operator and its Hermitian conjugate. 
We show that the bi-orthogonal tight-binding model well reproduces the dispersion relations of the continuum model when the complex scalar potential is sufficiently large.
\end{abstract}

\maketitle

\section{Introduction}
\label{sec:introduction}
The behavior of spatially periodic systems are determined by band structures and corresponding Bloch functions. 
Typically, bands are separated by gaps and oscillating modes with frequencies inside the gaps are prohibited in the periodic media. 
There are roughly two regimes; one is the nearly-free regime where the periodic potential is weak and can be treated as a perturbation to the free particle, and the other is the tight-binding regime where band structures are well approximated by lattice models based on localized basis functions. 
Closed periodic systems described by Hermitian Hamiltonians have been extensively studied and the behavior in both regimes is well understood, such as how band gaps open with weak periodic potentials and the relation between continuum models and tight-binding models \cite{ashcroft1976solid,
marzani1997maximally,
souza2001maximally,
marzani2012maximally}. 

Meanwhile, in open systems effectively described by non-Hermitian operators, intriguing phenomena which have no counterpart in closed systems have been revealed \cite{ashida2020non}, such as $\mathcal{PT}$ symmetry breaking \cite{bender1998real,
makris2008beam,
guo2009observation,
zheng2010pt,
ruter2010observation,
miri2012large,
regensburger2012parity,
chtchelkatchev2012stimulation,
schomerus2013topologically,
feng2014single,
hodaei2014parity,
peng2014parity,
peng2014loss,
poli2015selective,
zeuner2015observation,
mochizuki2016explicit,
ashida2017parity,
kawabata2017information,
xiao2017observation,
kawabata2018parity,
konotop2018odd,
li2019observation,
longhi2019non,
okugawa2019topological,
xiao2019observation,
mochizuki2020bulk,
mochizuki2020statistical,
wang2021observation,
acharya2022localization}, novel topological phenomena \cite{rudner2009topological,
esaki2011edge,
hu2011absence,
leykam2017edge,
weimann2017topologically,
xiao2017observation,
alvarez2018non,
shen2018topological,
gong2018topological,
kunst2018biorthogonal,
qi2018defect,
lieu2018topological,
dangel2018topological,
kunst2018biorthogonal,
yao2018edge,
ghatak2019new,
kawabata2019symmetry,
longhi2019topological,
sone2019anomalous,
yokomizo2019non,
borgnia2020non,
lieu2020tenfold,
okuma2020topological,
mochizuki2020bulk,
sone2020exceptional,
zhang2020correspondence,
zhang2020non,
bergholtz2021exceptional,
pan2021point,
mochizuki2021fate,
acharya2022localization}, and the emergence of exceptional points \cite{berry2004physics,
dembowski2004encircling,
heiss2012physics,
gao2015observation,
doppler2016dynamically,
chen2017exceptional,
kawabata2019classification,
miri2019exceptional,
okugawa2019topological,
zhang2019experimental,
sone2020exceptional,
zhang2020non,
hamazaki2021exceptional}, to name a few. 
Regarding huge progress on non-Hermitian topological phases made recently, which is a relatively new arena compared to other realms, the majority of studies are based on tight-binding models and continuum models are rarely explored \cite{longhi2021non,yokomizo2021non}. 
In addition, the relation between continuum models and tight-binding models in open systems has not been well understood in comparison to closed systems.

In the present work, we explore band structures of non-Hermitian continuum models in both nearly-free and tight-binding regimes.
In contrast to systems described by Hermitian Hamiltonians where infinitesimally small real potentials open gaps \cite{ashcroft1976solid}, we reveal that weak imaginary scalar potentials in open systems do not open band gaps but lead to the formation of exceptional points as long as the strength of the potential is smaller than a threshold value. 
We also show that imaginary vector potentials hinder the separation of low-energy bands when the scalar potential is weak, while they generate a different type of gaps referred to as point gaps~\cite{kawabata2019symmetry} when the scalar potential is strong, which is a unique feature of open systems. 
Furthermore, we obtain bi-orthogonal Wannier functions and construct tight-binding models based on Bloch wavefunctions of the non-Hermitian operator and its Hermitian conjugate. 
We show that the dispersion relations of the continuum model with complex scalar and/or vector potentials are well reproduced by the bi-orthogonal tight-binding model when the scalar potential is strong. 

The rest of this paper is organized as follows.
In Sec.~\ref{sec:setup}, we present our setup, which is a one-dimensional system with complex periodic scalar and vector potentials.
In Sec. \ref{sec:nearly-free}, we study the band structures in the nearly-free regime with small potentials. 
We find several unique behaviors with no counterpart in closed Hermitian systems; the emergence of exceptional points, the parameter dependence of gap sizes, and ring-shaped band structures.
In Sec. \ref{sec:tight-binding}, we explore the tight-binding regime where the periodic potential is large, and construct lattice models which well reproduce the band structures of the continuum models. 
The construction of the tight-binding models is carried out by obtaining Wannier functions which are composed of  bi-orthogonal Bloch functions of the non-Hermitian operator and its Hermitian conjugate. 
Section \ref{sec:summary} is devoted to summary.

\section{Setup}
\label{sec:setup}

We consider one-dimensional systems in the presence of periodic complex scalar and vector potentials, described by the wavefunction $\psi (x,t)$, which is a function of position $x$ and time $t$, obeying the Schr\"odinger equation
\begin{align}
    i\frac{\partial}{\partial t}\psi (x,t) = H_x \psi (x,t), 
    \label{eq:schroedinger-equation}
\end{align}
with the non-Hermitian operator
\begin{align}
    H_x = \frac{1}{2M}\left[-i\frac{\partial}{\partial x}-A(x)\right]^2+V(x),
    \label{eq:hamiltonian}
\end{align}
which we refer to as the non-Hermitian Hamiltonian. 
Here, $M=M^\ast$ is the mass, $A(x) \neq A^\ast (x)$ is the vector potential, and $V(x) \neq V^\ast(x)$ is the scalar potential.  
Note that, although we use terminologies in quantum mechanics, Eq. (\ref{eq:schroedinger-equation}) appears in various situations not restricted to quantum systems. 
Physical meanings of the wavefunction $\psi (x,t)$ and the Hamiltonian $H_x$ as well as various terms inside $H_x$ depend on the specific realizations of the non-Hermitian Schr\"odinger equation (\ref{eq:schroedinger-equation}).
For instance, such a non-Hermitian Hamiltonian phenomenologically describes the dynamics of electric fields inside materials with complex refractive indices \cite{makris2008beam,
guo2009observation,
yokomizo2021non}. 
For various other systems where non-Hermitian Hamiltonians emerge, we refer to review papers such as Refs~\cite{bender2007making,ashida2020non,bergholtz2021exceptional}. 
The complex vector and scalar potentials obey the same periodicity with a period $a$:
\begin{align}
    A(x+a) &= A(x),& V(x+a) &= V(x).
    \label{eq:periodic-potential}
\end{align}
We take the periodic boundary condition with a system size of $L=Na$, where $N$ is an integer. 
Throughout this paper, we take $a$ as the unit of length and $1/(2Ma^2)$ as the unit of complex energy, and we simply set $a=1$ and $1/(2Ma^2) = 1$.
In this paper, we explore the band structures of $H_x$ under various strength of the scalar and vector potentials.
We first note that, for the vector potential $A(x)$, we only need to study the case where $A(x)$ is constant and purely imaginary. 
This is because $A(x)$ can be expanded in Fourier series as $A(x)=A(x+1)=A+\sum_{l\neq0}A_l\exp(i2\pi lx)$, and the oscillating components other than the constant part can be gauged away by the transformation of the wavefunction $\psi (x)\rightarrow\psi (x)\exp\left(-i\int^x_0[A(x^\prime)-A\,]dx^\prime\right)$, and the eigenenergies are invariant under the transformation. 
Since a constant real vector potential just shifts the origin of the quasimomentum, we only need to consider a constant imaginary vector potential, as long as we are concerned with the energy band structure.

\begin{figure*}
    \centering
    \includegraphics[width=18cm]{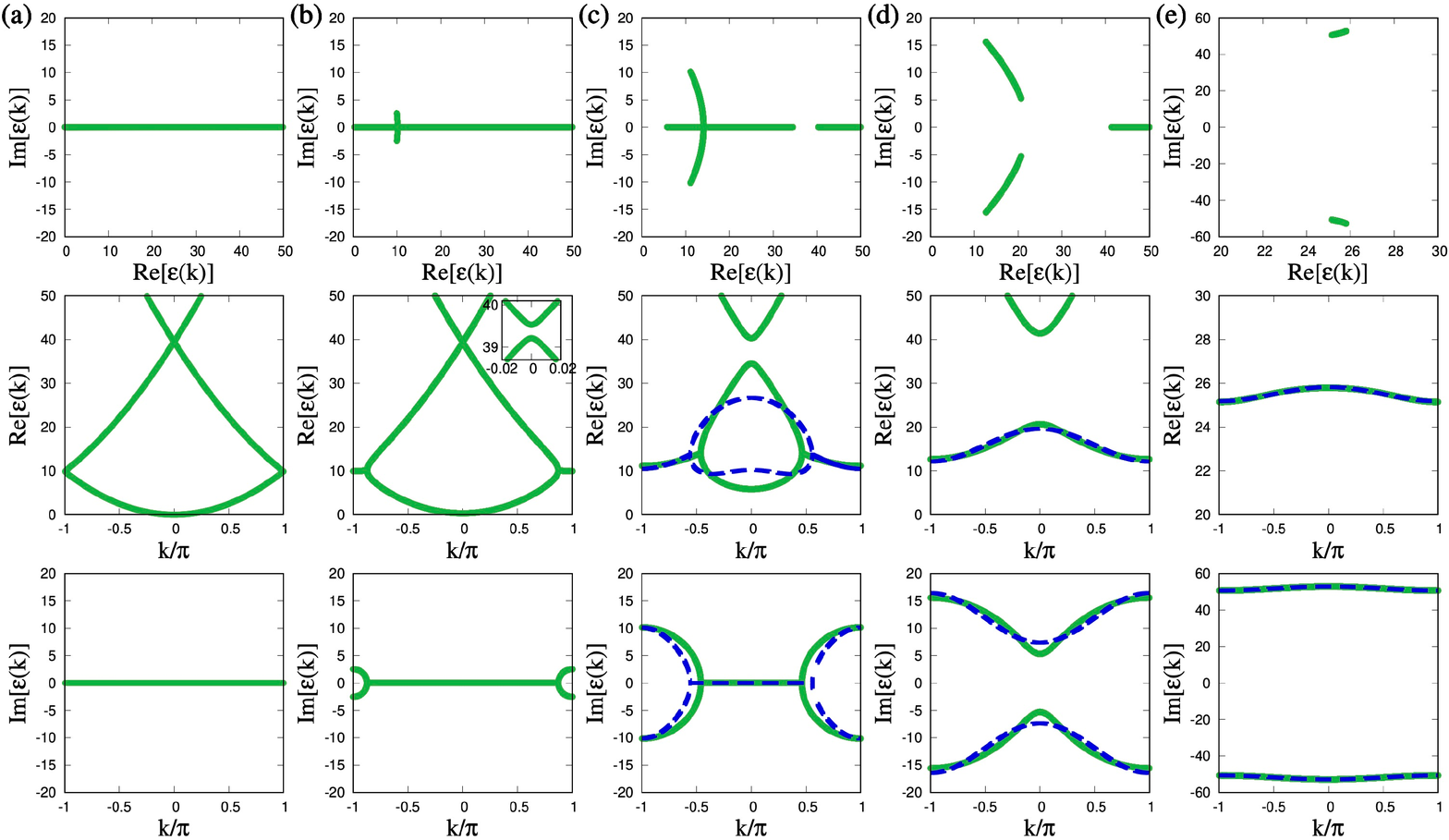}
    \caption{Eigenvalues in the complex plane and dispersion relations with $A=0$ and $V(x) = c \sin (2\pi x)$ (a) $c=0$, (b) $c=5i$, (c) $c=20i$, (d) $c=30i$, and (e) $c=80i$. 
    Green lines are obtained from the continuum model by solving Eq. (\ref{eq:eigen-equation_k}), and blue dashed lines in the center and bottom rows of (c)-(e) represent dispersion relations of first and second bands obtained from the bi-orthogonal tight-binding model discussed in Sec.~\ref{sec:tight-binding}. 
    In the bottom of (b)-(e), only Im$[\varepsilon_1(k)]$ and Im$[\varepsilon_2(k)]$ are shown but $\mathrm{Im}[\varepsilon_3(k)] = 0$ is not shown to avoid the figures becoming too crowded, while the third band is also shown in the top and center figures of (b)-(d). 
    In (d) and (e), Re$[\varepsilon_1(k)]$ and Re$[\varepsilon_2(k)]$ take the same value in the whole Brillouin zone.}
    \label{fig:eigenvalue_dispersion_pt}
\end{figure*}

The Bloch theorem for wavefunctions in a periodic potential holds also for non-Hermitian Hamiltonians. Namely, the eigenvalues and eigenstates are labeled by the band index $n$ and a quasimomentum $k=2\pi m/L$ with $m=1,2,\cdots,N$, obeying the eigenvalue equation
\begin{align}
    H_x \psi_k^n(x)=\varepsilon_n(k)\psi_k^n(x),
    \label{eq:eigen-equation_x}
\end{align}
where the eigenstate can be written as
\begin{align}
    \psi_k^n(x)=e^{ikx}u_k^n(x),
    \label{eq:bloch-function}
\end{align}
which we refer to as the Bloch state,
and $u_k^n (x)$ obeys the periodicity $u_k^n(x+1) = u_k^n(x)$. We note that since the Hamiltonian is non-Hermitian, the eigenvalue $\varepsilon_n (k)$ is generally a complex-valued function.
By defining $H_k = e^{-ikx}H_x e^{ikx}$, the eigenvalue equation becomes
\begin{align}
    H_k u_n^k(x)=\varepsilon_n(k) u_n^k(x).
    \label{eq:eigen-equation_k}
\end{align}
For convenience, we order bands $n = 1, 2, 3, \cdots$ in the following manner.
We define bands so that $\varepsilon_n(k)$ is a continuous function of $k$, and if bands do not contain exceptional points, we first order them according to the real parts and then order according to imaginary parts. 
To be more precise, we first take $\mathrm{min}\left( \text{Re}[\varepsilon_n(k)] \right) \leq \mathrm{min}\left( \text{Re}[\varepsilon_{n+1}(k)] \right)$, and if the minimum of the real part of two (or more) bands are the same, we take $\mathrm{min}\left( \text{Im}[\varepsilon_n(k)] \right) \leq \mathrm{min}\left( \text{Im}[\varepsilon_{n+1}(k)] \right)$. 
We will define later how to order bands which are mixed with exceptional points when discussing Fig.~\ref{fig:eigenvalue_dispersion_pt}.

For the complex scalar potential $V(x)$, we will mostly consider a sinusoidal potential
\begin{align}
    V(x) = c \sin (2\pi x)
    \label{eq:imaginary-potential}
\end{align}
with complex $c$, unless otherwise stated. 
In Fig.~\ref{fig:eigenvalue_dispersion_pt}, we plot the energy eigenvalues for various values of purely imaginary $c$ with increasing $|c|$ in the absence of the vector potential. 
We observe that the evolution of the band structure as $|c|$ increases is quite different from the textbook example of a periodic Hermitian potential. 
We will provide a quantitative understanding of the band structure both in the nearly-free and the tight-binding regimes.

Before proceeding to the next section, we briefly describe how we can numerically calculate the band structures such as those in Fig.~\ref{fig:eigenvalue_dispersion_pt}. 
We note that the method is the standard one often used for Hermitian Hamiltonians.
Noticing that $u_n^k(x)$ is periodic with period $1$, the eigenvalue equation, Eq.~(\ref{eq:eigen-equation_k}), can be expanded in Fourier series. 
Expanding $u_k^n(x)$ and $V(x)$ as $u_k^n(x) = \sum_{l} u^n_l (k)\exp(i2\pi lx)$ and $V(x)=\sum_lV_l\exp(i2\pi lx)$, Eq.~(\ref{eq:eigen-equation_k}) becomes a matrix equation for a given value of $k$
\begin{align}
    \sum_m  H_{lm}(k) u^n_m (k) = \varepsilon_n(k) u_l^n(k),
    \label{eq:blochmatrix}
\end{align}
where $H_{lm}(k)=(k+2\pi m-A)^2\delta_{lm}+V_{l-m}$.
When $V(x) = c \sin (2\pi x)$, only nonzero components of $V_l$ are $V_{-1} = ic/2$ and $V_1 = -ic/2$.
By truncating the Fourier series including an enough number of Fourier components, the matrix equation can be numerically diagonalized to obtain the eigenvalues and eigenvectors. 
Throughout the paper, we take $-40\leq l \leq 40$ for numerical calculations, with which we have confirmed to obtain the convergence in the calculation of the eigenvalues.

Recent studies on application of the band theory to non-Hermitian continuous models have mostly focused on the effect of the imaginary vector (gauge) potential and the resulting skin effects in continuous systems~ \cite{longhi2021non,yokomizo2021non}, whereas our work is more intended to elucidate the role of the imaginary scalar potential. 
There have also been studies, in the context of $\mathcal{PT}$-symmetric optics, on the band structure of continuous models under $\mathcal{PT}$-symmetric potentials~\cite{makris2008beam,guo2009observation,ruter2010observation}, whose analysis is, in our terminology, restricted to the nearly-free regime without a vector potential.
In these earlier studies on $\mathcal{PT}$-symmetric optics, the emergence of exceptional points has been discussed. 
In this paper we will give a coherent description connecting the nearly-free and the tight-binding regimes and qualitatively explain the emergence of exceptional points from the viewpoint of the bi-orthogonal tight-binding model. 

\section{Nearly-free regime}
\label{sec:nearly-free}
We start from the nearly-free regime, where the strength of the potential is relatively small, so that $V(x)$ can be regarded as a perturbation from a free case $V(x) = 0$.

\subsection{Without a vector potential}
\label{subsec:nearly-free_without-A}

We first analyze the situation where $c$ is purely imaginary and there is no vector potential. 
The corresponding band structures are given in Fig.~\ref{fig:eigenvalue_dispersion_pt}. 
When there is no potential, the first and the second bands touch at $k = \pm \pi$, and the second and the third bands touch at $k = 0$ as shown in Fig.~\ref{fig:eigenvalue_dispersion_pt} (a), which is well known from the Hermitian band theory. 
As we add a small imaginary potential $c \neq 0$, we observe drastically different behaviors for the band touching points at $k = \pm \pi$ and $k = 0$, as shown in Fig.~\ref{fig:eigenvalue_dispersion_pt} (b). 
In this case, exceptional points appear, and therefore we order first and second bands such that $\mathrm{Im}[\varepsilon_1(k)]\leq\mathrm{Im}[\varepsilon_{2}(k)]$ ($\mathrm{Re}[\varepsilon_1(k)]\leq\mathrm{Re}[\varepsilon_{2}(k)]$) is satisfied in the region where the real (imaginary) parts of eigenvalues are degenerate.
The real parts of $\varepsilon_1(k)$ and $\varepsilon_2(k)$ around $k = \pm \pi$ form a degenerate line while the imaginary parts open a gap. There also appear exceptional points near $k = \pm \pi$ where the eigenvalues coalesce. 
On the other hand, regarding the degeneracy at $k=0$ in the absence of the potential, a real gap is opened and thus $\varepsilon_2(k)$ and $\varepsilon_3(k)$ are separated, similar to what happens in a Hermitian potential.
We note that a similar exceptional point structure has been found for a $\mathcal{PT}$-symmetric system in Ref.~\cite{makris2008beam}, where they employ a different form of a scalar potential; we thus expect that such a formation of exceptional points is a generic feature of systems under non-Hermitian $\mathcal{PT}$ symmetric periodic potentials, and the analysis below can also be applied to other forms of potentials, \textit{mutatis mutandis}, to describe the exceptional points.

\begin{figure}
    \centering
    \includegraphics[width=9cm]{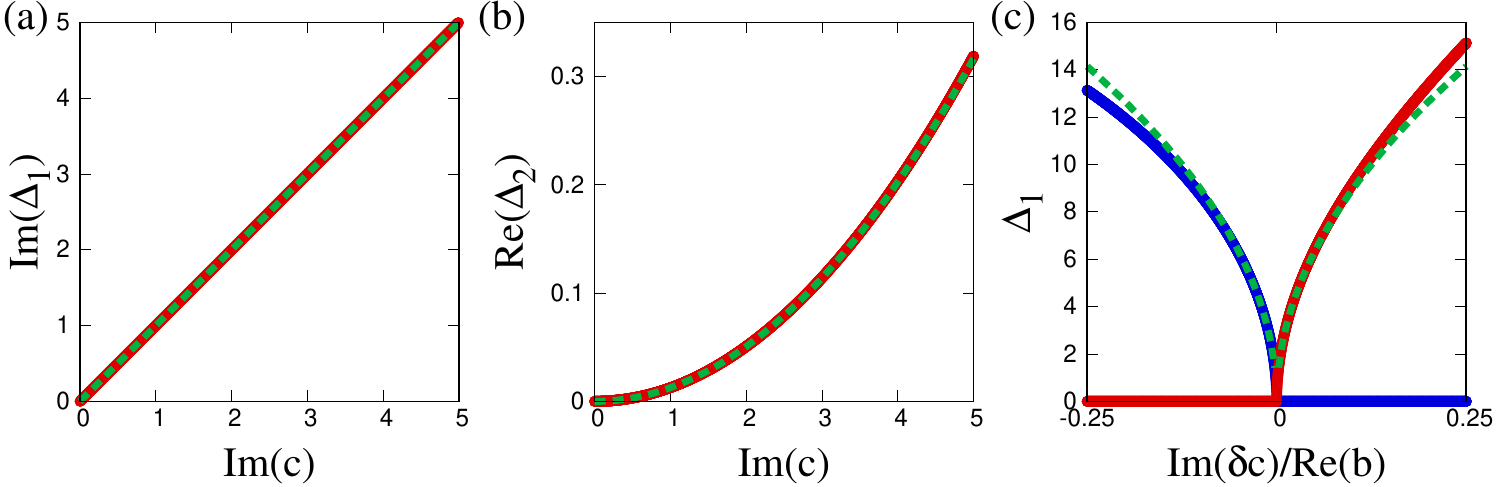}
    \caption{The gap sizes $\Delta_1$ and $\Delta_2$ as functions of the strength of the scalar potential for $A = 0$. 
    The potentials are (a),(b) $V(x) = c \sin (2\pi x)$ with purely imaginary $c$ and (c) $V(x) = b\cos (2\pi x) + c \sin (2\pi x)$ with $b=20$ and purely imaginary $c$. 
    Solid (red and blue) lines show numerical results and dashed (green) lines correspond to the results from the perturbation analysis, (a) $\Delta_1=c$, (b) $\Delta_2=|c|^2 /8\pi^2$, and (c) $\Delta_1=\sqrt{\pm2b|\delta c|}$.
    In (c), red and blue lines correspond to $\mathrm{Re}(\Delta_1)$ and $\mathrm{Im}(\Delta_1)$, respectively.}
    \label{fig:gap-size}
\end{figure}

The behavior at $k = \pm \pi$ between the first and second bands can be understood from a simple first-order perturbation theory. Focusing on the band degeneracy at $k = \pi$, the periodic parts for Bloch states of the first and the second bands before adding the potential $V(x)$ are simply $u^1_{k}(x) = 1$ and $u^2_{k}(x) = e^{-i2\pi x}$.
Considering the perturbation theory taking $u^1_{k}(x)$ and $u^2_{k}(x)$ as non-perturbative states is equivalent to considering only $l = 0$ and $l=-1$ terms in the matrix equation (\ref{eq:blochmatrix}) of the Fourier-transformed eigenvalue equation. 
Therefore, considering the potential $V(x) = c \sin (2\pi x)$ as a perturbation, the matrix elements of the Hamiltonian with respect to $u^1_{k}(x)$ and $u^2_{k}(x)$ are
\begin{align}
    \begin{pmatrix}
    H_{-1,-1} &
    H_{-1,0} \\
    H_{0,-1} &
    H_{0,0}
    \end{pmatrix}
    =
    \begin{pmatrix}
    (k-2\pi)^2 & ic/2 \\
    -ic/2 & k^2
    \end{pmatrix},
    \label{eq:firstgap}
\end{align}
The energy gap at $k = \pi$ is then determined by the eigenvalues of the above matrix, which are $\pi^2 \pm c/2$. When $c$ is purely imaginary, the gap will thus be purely imaginary as we numerically observe in Fig.~\ref{fig:eigenvalue_dispersion_pt} (b). 
As shown in Fig. \ref{fig:gap-size} (a), the numerically obtained gap size
\begin{align}
    \Delta_1=\varepsilon_2(\pi)-\varepsilon_1(\pi)
    \label{eq:gap-size_1}
\end{align}
agrees well with the analytical result of the first-order perturbation theory. 
Figure \ref{fig:eigenvalue_dispersion_pt} (c) shows that the two exceptional points, which emerge from $k = \pm \pi$, approach toward $k = 0$ as $|c|$ is increased. 
These exceptional points collide at a threshold value $|c|$, which we find to be around $|c| \approx 29$, and the two bands are separated after the collision as shown in Fig. \ref{fig:eigenvalue_dispersion_pt} (d). 
The behavior explained above is unique to open systems described by non-Hermitian Hamiltonians since real (Hermitian) periodic potentials with infinitesimally weak strength separate the bands in closed systems \cite{ashcroft1976solid}. 
The strength for the imaginary part of the complex potential adopted in Ref. \cite{makris2008beam} is around $c \approx 20$ in the terminology of our paper, and thus the regime we explore in this paper is of relevance to $\mathcal{PT}$-symmetric optical systems.

The gap at $k=0$ between the second and the third bands behave differently.
Non-perturbative states which are degenerate at $k = 0$ are components with $l = \pm 1$ in Eq. (\ref{eq:blochmatrix}). 
However, matrix elements of the periodic potentials are all zero between these two states, $V_{\pm 2} = 0$, and thus we need to consider higher order terms. We can include a higher order term by including also the first band into the calculation. 
Thus considering $l = -1$, 0, 1 components, the matrix elements of the Hamiltonian are
\begin{align}
    \begin{pmatrix}
    H_{-1,-1} & H_{-1,0} & H_{-1,1} \\
    H_{0,-1} & H_{0,0} & H_{0,1} \\
    H_{1,-1} & H_{1,0} & H_{1,1}
    \end{pmatrix}
    =
    \begin{pmatrix}
    (k-2\pi)^2 & ic/2 & 0 \\
    -ic/2 & k^2 & ic/2 \\
    0 & -ic/2 & (k+2\pi)^2
    \end{pmatrix}.
\end{align}
The eigenvalues of this matrix at $k = 0$ are $4\pi^2$ and $\frac{1}{2}\left( 4\pi^2 \pm \sqrt{(4\pi^2)^2 + 2c^2}\right) \approx -\frac{c^2}{8\pi^2},$ $4\pi^2+\frac{c^2}{8\pi^2}$. 
Thus, the size of the gap between the second and the third gap at $k=0$,
\begin{align}
    \Delta_2=\varepsilon_3(0)-\varepsilon_2(0),
    \label{eq:gap-size_2}
\end{align}
is $-c^2/8\pi^2$ which is a real number when $c$ is purely imaginary. 
In Fig. \ref{fig:gap-size} (b), we plot the numerically obtained gap size as well as the analytical expression from the perturbation theory, and we find an almost perfect agreement.

\begin{figure}
    \centering
    \includegraphics[width=9cm]{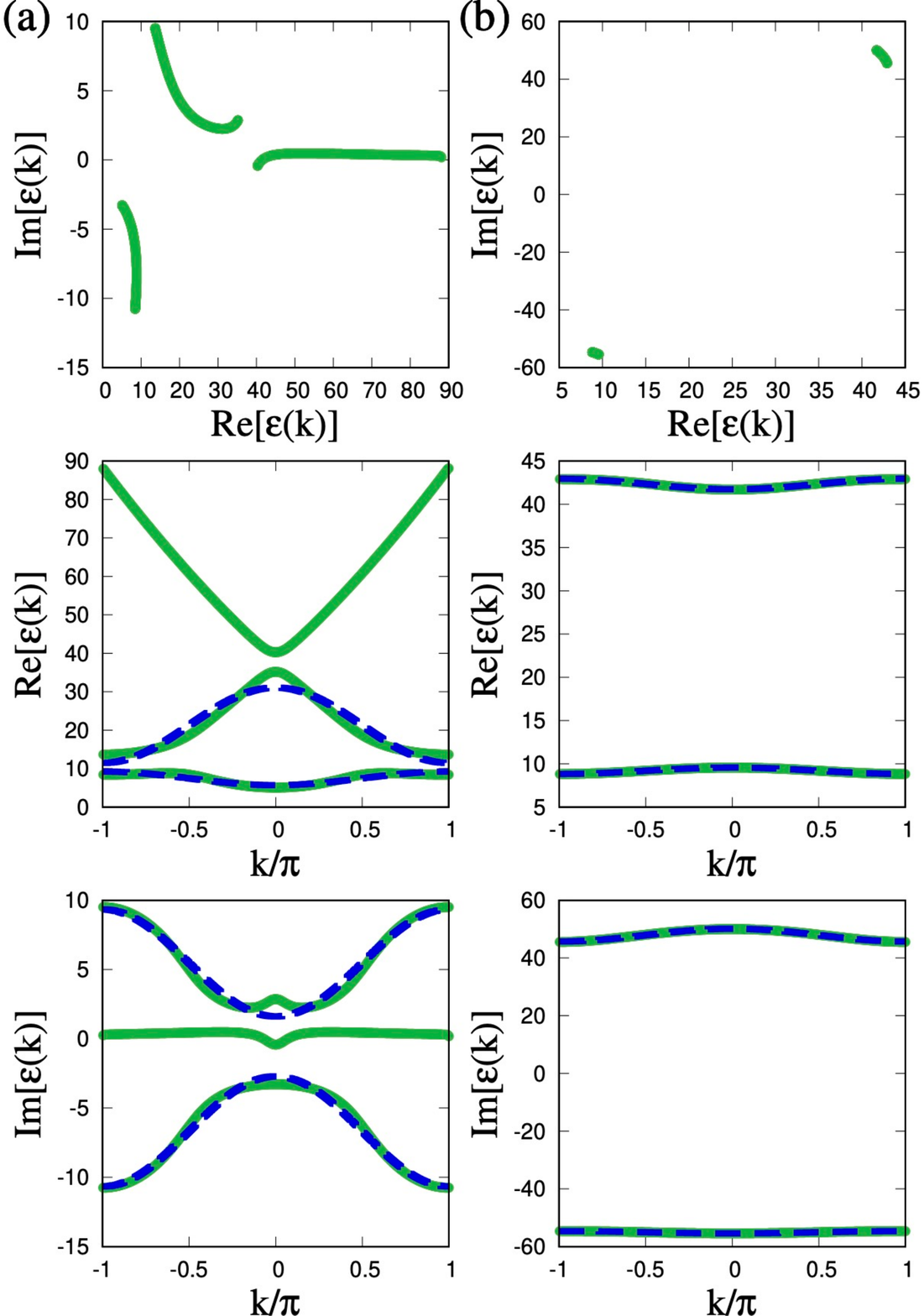}
    \caption{Eigenvalues in the complex plane and dispersion relations when $A=0$ with $V(x) = c \sin (2\pi x)$ when (a) $c=5+20i$ and (b) $c=20+80i$. 
    The green lines are calculated from the continuous model, and the blue dashed lines are the results from tight-binding approximation.}
    \label{fig:eigenvalue_dispersion_without-A}
\end{figure}

In Fig. \ref{fig:eigenvalue_dispersion_without-A} (a), we show band structures when $c$ has both real and imaginary parts. 
In such a case, the gaps also have real and imaginary parts, as we also see from the perturbation results above, and the lowest two bands do not touch anywhere in the Brillouin zone.

We note that the difference in the gap opening at $k = \pm \pi$ and $k = 0$ is not just the gap sizes being proportional to $c$ or $c^2$. 
As we have already clarified, for the gap at $k = \pm\pi$, the exceptional points between the first and second bands emerge, and there is no point or line gap opening in the complex energy plane with imaginary $c$. 
On the other hand, the gap at $k = 0$ does not lead to exceptional points, and there is a line gap in the complex plane, similar to the gap opening in the real (Hermitian) periodic potentials.

\begin{figure}[tbp]
    \centering
    \includegraphics[width=9cm]{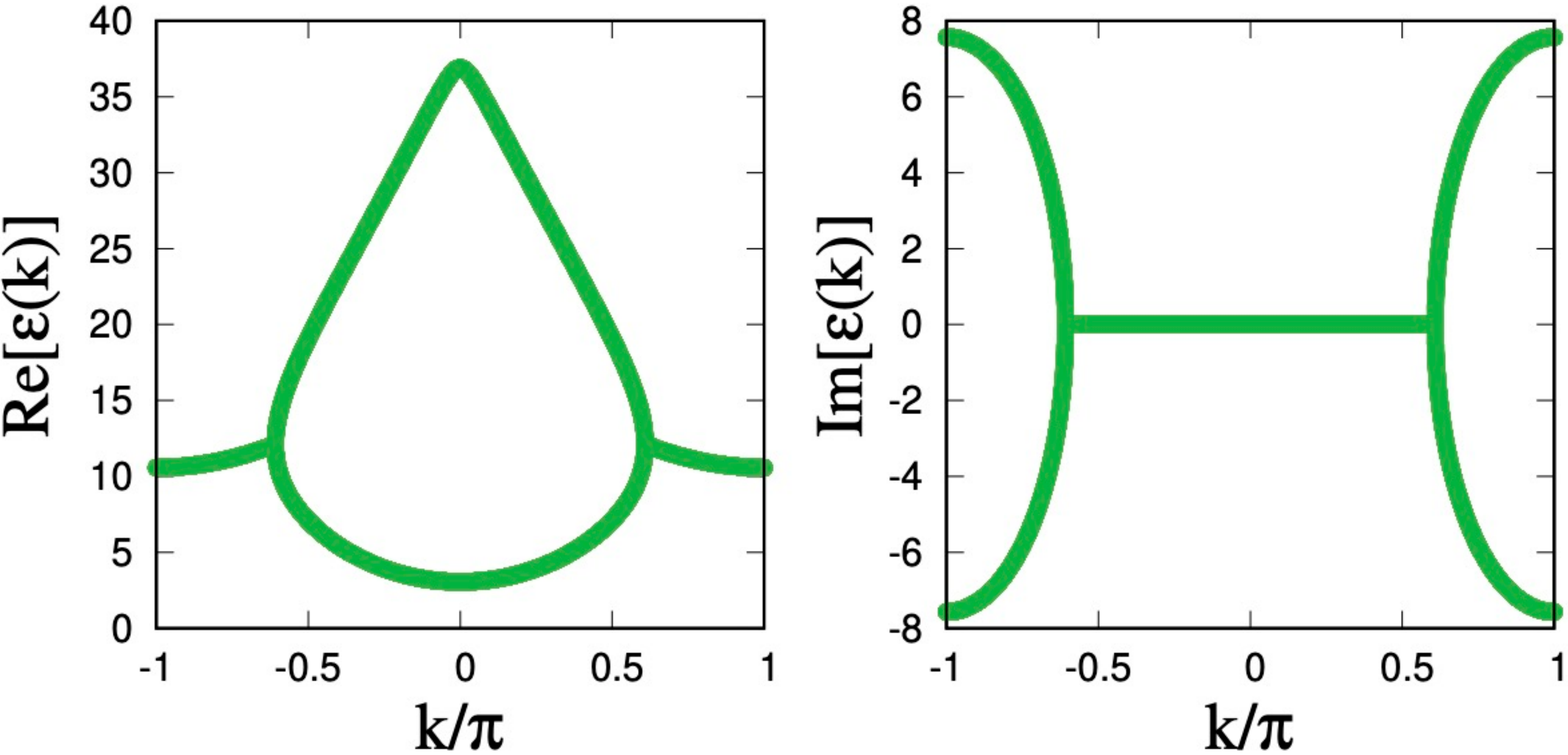}
    \caption{The dispersion relation when $A=0$ and $V(x) = b\cos (2\pi x) + c \sin (2\pi x)$ with $b = 20$ and $c=25i$.}
    \label{fig:dispersion_near-ep}
\end{figure}

One may notice that the imaginary parts of the dispersion relations are symmetric around $\mathrm{Im}[\varepsilon (k)] = 0$ in Fig.~\ref{fig:eigenvalue_dispersion_pt}. 
This is a direct consequence of the $\mathcal{PT}$ symmetry of the Hamiltonian $H_{-x}^\ast=H_x$ where the purely imaginary character of $V(x) = c \sin (2\pi x)$ plays a crucial role here. 
In the eigenvalue equation of the $n$-th band, $H_x \psi_k^n (x) = \varepsilon_n (k) \psi_k^n (x)$, taking its complex conjugation and transforming $x \to -x$, one obtains $H_x [\psi_k^n (-x)]^* = \varepsilon_n^\ast (k) [\psi_k^n (-x)]^*$ from the $\mathcal{PT}$ symmetry. 
This relation shows that if there exists an eigenvalue $\varepsilon_n (k)$ with nonzero imaginary part, its complex conjugate $\varepsilon_n^\ast (k)$ should also be an eigenvalue. 
Thus, the eigenvalues should appear either purely real or appear in complex conjugate pairs, which explains the symmetry around $\mathrm{Im}[\varepsilon (k)] = 0$. 
We note, in particular, that when the lowest two bands are separated, these two bands are the complex conjugate pairs obeying $\varepsilon_1 (k) = \varepsilon_2^\ast (k)$.

While we focus mainly on the potential of the form $V(x) = c\sin(2\pi x)$ in this paper, there are also many other types of periodic potentials which have the periodicity of $x \to x + 1$. 
Covering general shapes of the periodic potential is beyond the scope of the present paper. 
However, before proceeding to add a vector potential, we want to mention one specific case $V(x) = b \cos (2\pi x) + c \sin (2\pi x)$, where $b$ is real and $c$ is imaginary, which shows a particularly noticeable feature related to the non-Hermiticity of the periodic potential. 
When $ib = c$, the scalar potential takes the form $V(x) = b e^{i2\pi x}$, and therefore its Fourier component in the eigenvalue equation (\ref{eq:blochmatrix}) only has one nonzero component $V_1 = b$. 
The matrix $H_k$ then takes the lower triangular form with diagonal elements taking $H_{ll} = (k+2\pi l)^2$. 
Therefore, the eigenvalues are $k^2$, properly folded in the Brillouin zone, which are exactly the same as the eigenvalues in the absence of the vector and scalar potentials. 
Even though the eigenvalues for $V(x) = 0$ and $V(x) = b e^{i2\pi x}$ are the same, their responses to external perturbations are very different. 
We consider adding $\delta c \sin (2\pi x)$ with purely imaginary $\delta c$ to the potential $V(x)$. 
For $b=0$, $V(x)=\delta c \sin (2\pi x)$ is nothing but the situation treated above and the imaginary gap $\delta c$ opens at $k = \pi$ between the first and the second band with the formation of exceptional points. 
On the other hand, when we add $\delta c \sin (2\pi x)$ to $V(x) = b e^{i2\pi x}$, the matrix elements of $H_k$ for the lowest two bands around $k = \pi$ are
\begin{align}
    \begin{pmatrix}
    H_{-1,-1} &
    H_{-1,0} \\
    H_{0,-1} &
    H_{0,0}
    \end{pmatrix}
    =
    \begin{pmatrix}
    (k-2\pi)^2 & i\delta c/2 \\
    b -i\delta c /2 & k^2
    \end{pmatrix}.
\end{align}
Note that when $\delta c = 0$, this truncated $2\times2$ matrix is essentially the Jordan normal form at $k = \pi$ and therefore $k = \pi$ is an exceptional point.
The energy eigenvalues at $k = \pi$ are then $\varepsilon_{1,2}(k=\pi) = \pi^2 \pm \sqrt{(b - i\delta c/2)(i\delta c/2)}$. 
When $|\delta c|$ is small, we thus obtain $\varepsilon_{1,2}(\pi) \approx \pi^2 \pm i\sqrt{b|\delta c|/2}$ when $\mathrm{Im}(\delta c) > 0$ and $\varepsilon_{1,2}(\pi) \approx \pi^2 \pm \sqrt{b|\delta c|/2}$ when $\mathrm{Im}(\delta c) < 0$. Thus, depending on the sign of $\mathrm{Im}(c)$, either a real or an imaginary gap opens, with a noticeable square-root dependence of the gap size $\propto \sqrt{|\delta c|}$. 
In Fig.~\ref{fig:dispersion_near-ep}, we plot the dispersion relation when $\delta c = 5i$. 
We indeed observe that an imaginary gap opens at $k = \pm \pi$, and its size shows an expected square-root behavior as shown in Fig. \ref{fig:gap-size} (c). This square-root sensitivity to the added perturbation is a characteristic feature of physics around exceptional points~\cite{heiss2012physics}.

\begin{figure*}
    \centering
    \includegraphics[width=18cm]{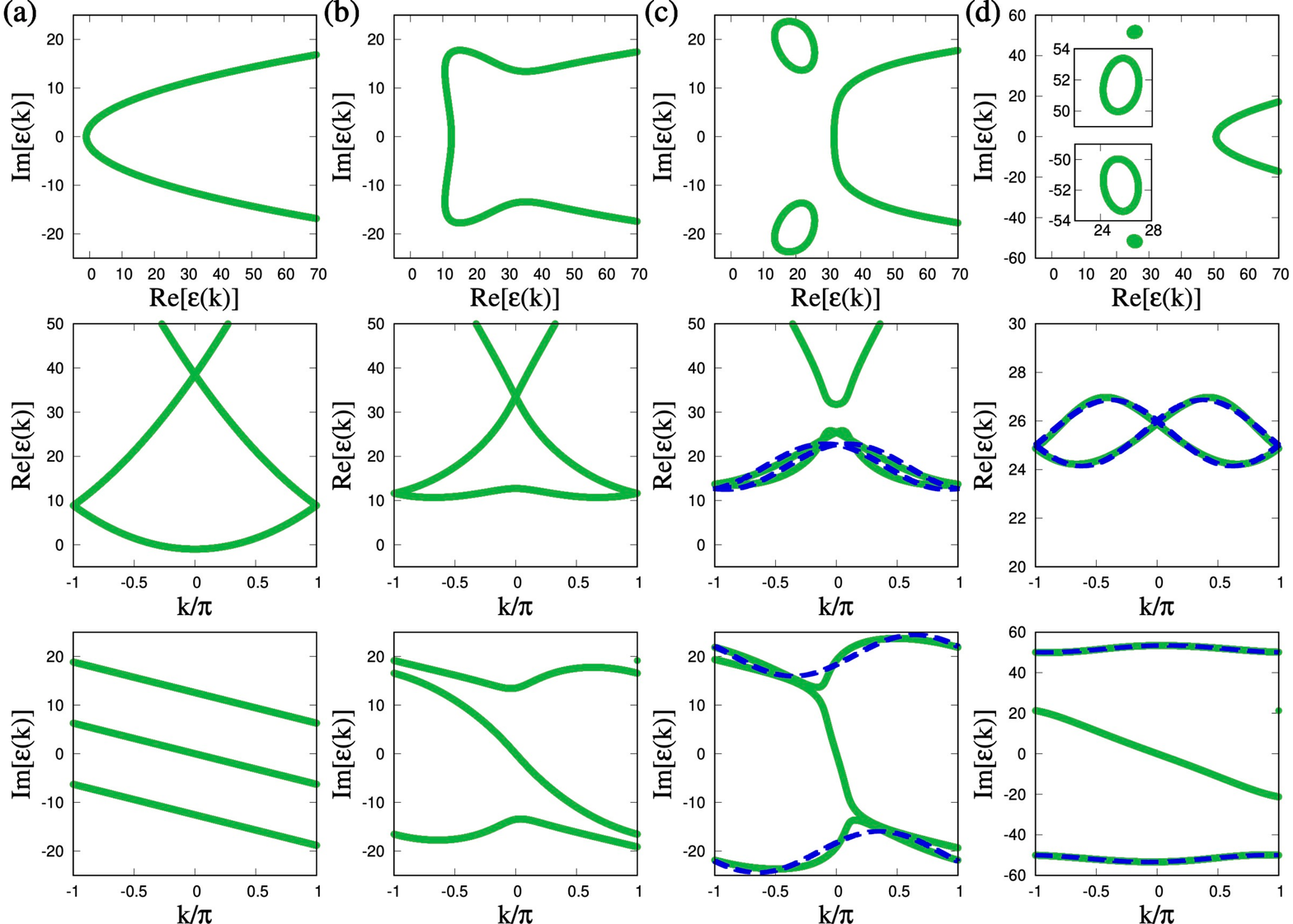}
    \caption{Eigenvalues in the complex plane and dispersion relations when $A=i$ and $V(x) = c \sin (2\pi x)$ with (a) $c=0$, (b) $c=30i$, (c) $c=40i$, and (d) $c=80i$. 
    The green lines are calculated from the continuous model, and the blue dashed lines are the results from tight-binding approximation.}
    \label{fig:eigenvalue_dispersion_with-A}
\end{figure*}

\subsection{With a vector potential}
\label{subsec:nearly-free_with-A}

When a constant imaginary vector potential $A$ is added, the dispersion in the absence of the scalar potential $V(x)$ becomes $\varepsilon(k) = k^2 + A^2 - 2Ak$, which should be properly folded when the first Brillouin zone is considered. While the real part of the energy is just shifted by a constant amount, $\mathrm{Re}[\varepsilon(k)] = k^2-[\mathrm{Im}(A)]^2$, the imaginary part shows a linear dependence on $k$ as $\mathrm{Im}[\varepsilon(k)] = -2\mathrm{Im}(A)k$, which has an important consequence on the gap opening when a scalar potential is added.

Although we focus in this paper on the periodic boundary condition, we note that, in the presence of an imaginary vector potential, the non-Hermitian skin effect occurs under the open boundary condition, as discussed in Refs. \cite{longhi2021non, yokomizo2021non}. 
When we analyze physical properties which are not affected by the boundary condition, such as the dynamics of a wavepacket within a timescale where it does not reach the edge of the system \cite{mao2021boundary}, the analysis we give in this paper under the periodic boundary condition is experimentally relevant.
Upon studying wavepacket dynamics, one needs to make sure to construct wavepackets only from the lowest bands. If components from higher bands enter, these components may grow in time if they have large imaginary energy. In practical experiments, one needs to look for a right balance between the evolution time and the growth of unwanted components in the wavepacket.

In Fig.~\ref{fig:eigenvalue_dispersion_with-A}, we plot the energy dispersion in the presence of $A = i$ as we add a scalar potential $V(x) = c \sin (2\pi x)$ with a purely imaginary $c$. 
The first noticeable feature of adding an imaginary vector potential is that, in the absence of the scalar potential, the energy bands are not degenerate at any point in the Brillouin zone. 
For example, at $k = \pm \pi$, the real parts of the energies are degenerate between the first and the second bands, but their imaginary parts are different, and thus the first and the second bands are not degenerate in the complex plane. 
Because of this absence of the band degeneracy, adding a small periodic scalar potential does not lead to gap opening. 
As one increases the strength of $V(x)$, the degeneracy of the real parts at $k = \pm \pi$ between the first and the second bands and that at $k = 0$ between the second and the third bands persist, while the imaginary part of the first band approaches the imaginary parts of the second and third bands at $k = \pm \pi$ and $k = 0$, as shown in Fig. \ref{fig:eigenvalue_dispersion_with-A} (b). 
At a threshold value of $|c|$, the three bands become degenerate in the complex plane, leading to the gap opening as in Fig.~\ref{fig:eigenvalue_dispersion_with-A} (c). 
The separated two bands form closed circles in the complex plane, indicating the nontrivial point-gap topology of these separated bands \cite{gong2018topological,kawabata2019symmetry}.

\section{tight-binding models}
\label{sec:tight-binding}
As we have seen, when the strength of the periodic potential is increased, the two lowest energy bands separate. 
When the strength of the periodic potential is large enough, we can describe the separated bands in terms of the tight-binding approximation. 
As we shall see, unlike the case of Hermitian periodic potentials where orthogonal basis functions can be used for the tight-binding basis, the bi-orthogonal basis composed from Bloch wavefunctions of $H_x$ and $H_x^\dagger$ should be utilized for the tight-binding basis of a non-Hermitian Hamiltonian. 
We first discuss how we can construct the bi-orthogonal tight-binding basis, and then we apply the construction to our Hamiltonian.

\subsection{Definition of the bi-orthogonal basis}
\label{subsec:tight-binding_definition}
While eigenfunctions of Hermitian Hamiltonians with different eigenvalues are orthogonal, eigenfunctions of non-Hermitian Hamiltonians, such as Eq. (\ref{eq:eigen-equation_x}), are generally not orthogonal,
\begin{align}
    \int_{-L/2}^{L/2}dx[\psi_{k'}^{n'}(x)]^\ast\psi_k^n(x)\neq0
    \label{eq:not-orthogonal}
\end{align}
even when $k \neq k'$ or $n \neq n'$.
The Bloch wavefunctions thus do not give rise to a set of orthonormal basis states. 
Instead, it is useful to consider a bi-orthogonal basis set \cite{brody2013biorthogonal} utilizing eigenfunctions of $H_x^\dagger$, which is defined as an operator satisfying the relation $\int_{-L/2}^{L/2}dx\phi^\ast(x)H_x\psi(x)=\int_{-L/2}^{L/2}dx[H_x^\dagger\phi(x)]^\ast\psi(x)$ for any smooth functions with the periodic boundary conditions $\phi (x + L) = \phi(x)$ and $\psi(x + L) = \psi(x)$. 
It is easy to show that the explicit form of $H_x^\dagger$ is
\begin{align}
    H_x^\dagger=\left(-i\frac{\partial}{\partial x}-A^\ast\right)^2+V^\ast(x).
    \label{eq:hamiltonian_dagger}
\end{align}
Since $V^* (x)$ is again periodic with $x \to x + 1$, the Bloch theorem also holds and thus the eigenstates of $H_x^\dagger$ can again be labeled by the band index $n$ and the quasi-momentum $k$
\begin{align}
    H_x^\dagger\tilde{\psi}_k^n(x)=\tilde{\varepsilon}_n(k)\tilde{\psi}_k^n(x).
    \label{eq:eigen-equation_x_dagger}
\end{align}
Since the set of eigenvalues of $H_x^\dagger$ are the complex conjugates of $\{\varepsilon_n(k)\}$, we take the band indices of $\tilde{\varepsilon}_n(k)$ such that  $\tilde{\varepsilon}_n(k)=\varepsilon_n^\ast(k)$ is satisfied. 
We can easily show that $\{\tilde{\psi}_k^n(x)\}$ and $\{\psi_k^n(x)\}$ constitute the bi-orthogonal basis set,
\begin{align}
    \langle\tilde{\psi}_{k'}^{n'}|\psi_k^n\rangle=\int_{-L/2}^{L/2}dx[\tilde{\psi}_{k'}^{n'}(x)]^\ast\psi_k^n(x)=\delta_{nn'}\delta_{kk'},
    \label{eq:bi-orthogonal_bloch-function}
\end{align}
where we introduced the 'braket' notation, such as $\bra{\tilde{\psi}_k^n}=\int dx\bra{x}[\tilde{\psi}_k^n(x)]^\ast,\,\ket{\psi_k^n}=\int dx\psi_k^n(x)\ket{x}$, and $\langle x' | x \rangle = \delta(x-x')$ where the range of the integral is $-L/2 \leq x \leq L/2$. 
Using the bi-orthogonal Bloch eigenfunctions, we now proceed to define bi-orthogonal Wannier functions. 

For later convenience, we define Wannier functions involving multiple bands $n = 1, 2, \cdots$. 
When we want to construct Wannier functions from the Bloch wavefunctions $\psi_k^{n}(x)$ and $\tilde{\psi}_k^{n}(x)$, we can generally mix these bands using a unitary matrix $U(k)$ to define the Wannier functions by
\begin{align}
    w_n^m(x)&=\frac{1}{\sqrt{N}}\sum_{k,n'}e^{-ikm}U_{n'n}(k)\psi_k^{n'}(x),
    \label{eq:wannier-function}\\
    \tilde{w}_n^m(x)&=\frac{1}{\sqrt{N}}\sum_{k,n'}e^{-ikm}U_{n'n}(k)\tilde{\psi}_k^{n'}(x),
    \label{eq:wannier-function_tilde}
\end{align}
where the sum on $n^\prime$ is over the bands with which we want to construct Wannier functions.
We assume that exceptional points, where the number of eigenvectors reduce, appear only at most in discrete points in momentum space, which is the case relevant in the analysis of this paper. Generalization of the method to include scenarios where continuous exceptional lines can appear is left for future works.
The constructed Wannier functions are bi-orthogonal, 
\begin{align}
    \langle\tilde{w}_{n'}^{m'}|w_n^m\rangle
    =\delta_{nn'}\delta_{mm'}.
    \label{eq:bi-orthogonal_wannier-function}
\end{align}
By appropriately choosing the unitary matrix $U(k)$,
these Wannier functions can be spatially localized, as in the Hermitian case. 
We label the unit cells so that $w_n^{0}(x)$ and $\tilde{w}_n^{0}(x)$ are localized in the $0$-th unit cell. 
Then, since $w_n^m(x) = w_n^0 (x - m)$ and $\tilde{w}_n^m(x) = \tilde{w}_n^0 (x - m)$, $w_n^{m}(x)$ and $\tilde{w}_n^{m}(x)$ are localized in the $m$-th unit cell, and therefore we can use the states $w_n^{m}(x)$ and $\tilde{w}_n^{m}(x)$ to represent tight-binding basis states for sites within the $m$-th unit cell.
In the basis of the bi-orthogonal Wannier functions, we can write down the tight-binding model corresponding to our continuous Hamiltonian as
\begin{align}
    H_\text{t}=\sum_{n,n'=1}^{n_0}\sum_{m,m'=1}^N
    t^{m-m'}_{nn'}\ket{w_n^m}\bra{\tilde{w}_{n'}^{m'}},
    \label{eq:lattice-hamiltonian}
\end{align}
where $n_0$ is the number of bands which we include and the matrix elements are
\begin{align}
    t^{m-m'}_{nn'}&=\bra{\tilde{w}_n^m}H\ket{w_{n'}^{m'}}
    \nonumber\\
    &=\frac{1}{N}\sum_{kl}\varepsilon_l(k)U^\dagger_{nl}(k)U_{ln'}(k)e^{-ik(m-m')}.
    \label{eq:hopping}
\end{align}
The matrix elements with $m-m^\prime = 0$ represent intra-cell hoppings and on-site energies, whereas $m-m^\prime = \pm 1$ represent hoppings between adjacent cells.
The difference from Hermitian tight-binding models is that the non-Hermitian tight-binding models obtained by the procedure elucidated above are based on not orthogonal but bi-orthogonal Wannier functions $\{\tilde{w}_n^m(x)\}$ and $\{w_n^m(x)\}$.

\begin{figure}[tbp]
    \centering
    \includegraphics[width=\columnwidth]{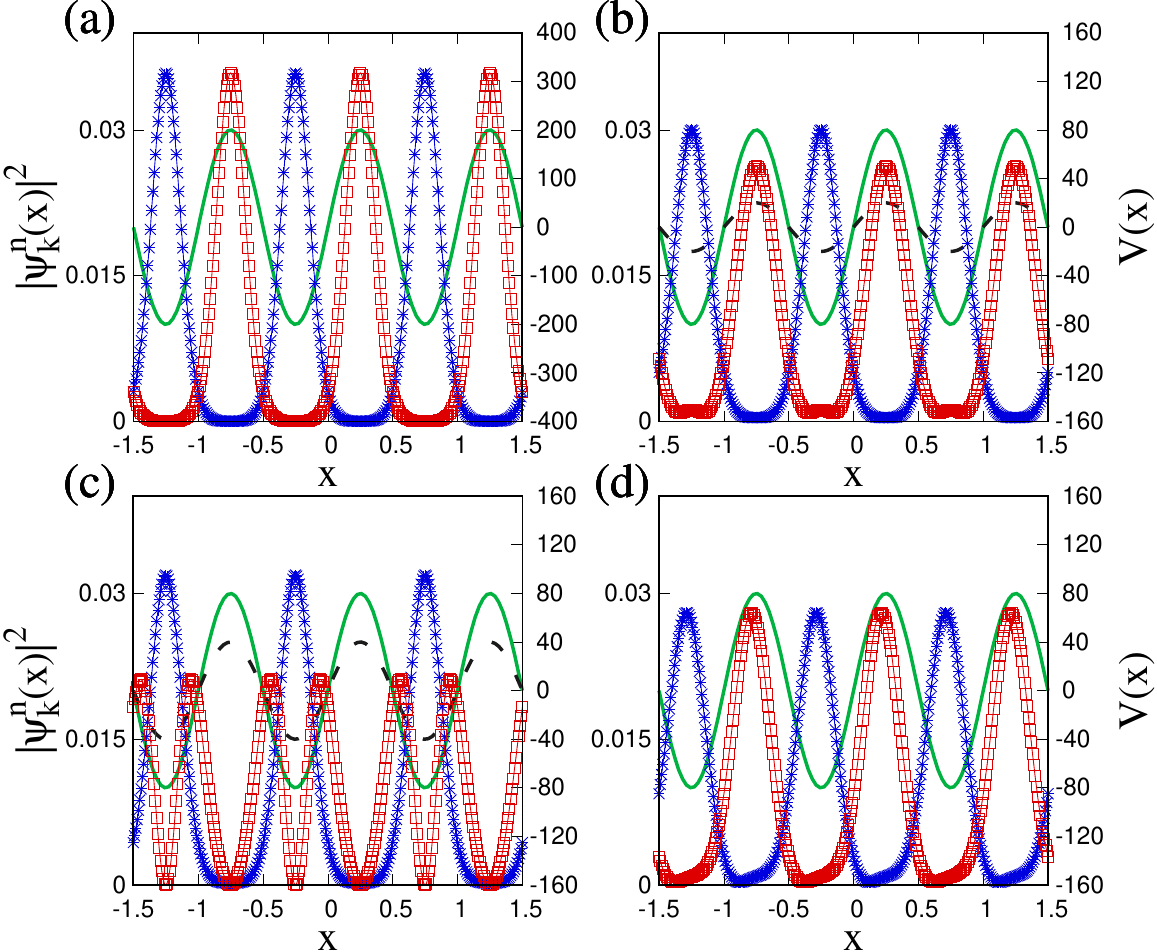}
    \caption{Bloch wavefunctions at $k=0$, $\psi_{k=0}^n (x)$, under  $V(x) = c \sin (2\pi x)$ when (a) $A=0,\,c=200i$, (b) $A=0,\,c=20+80i$, (c) $A=0,\,c=40+80i$, and (d) $A=1,\,c=80i$. 
    Red squares and blue asterisks respectively correspond to first and second bands $n=1,2$. 
    Green solid lines and black dashed lines are imaginary parts and real parts (if nonzero) of $V(x)$, respectively.}
    \label{fig:bloch-function}
\end{figure}

\subsection{When $V(x) = c \sin (2\pi x)$ is purely imaginary}
\label{subsec:imaginary-potential_without-A}

We first consider the situation where the periodic scalar potential is $V(x) = c\sin (2\pi x)$ with purely imaginary $c$ and no vector potential is present, $A = 0$. 

\subsubsection{Large $|c|$: Wannier functions constructed from individual bands}
\label{subsubsec:large-c}
When $|c|$ is large, as seen from Figs.~\ref{fig:eigenvalue_dispersion_pt} (d) and (e), the lowest two bands are separated in the complex plane. We first discuss that these two bands can be understood from the tight-binding approximation using Wannier functions constructed from each band separately.

We first plot the Bloch states $\psi_{k=0}^n (x)$ for the lowest two bands $n = 1,2$ with large $|c|$ in Fig.~\ref{fig:bloch-function} (a). 
We observe that the Bloch state of the first band $\psi_{k=0}^1(x)$ is localized at the minimum of the imaginary part of the scalar potential Im[$V(x)$] whereas that of the second band $\psi_{k=0}^2(x)$ is localized at the maximum of Im[$V(x)$].
We have confirmed that, with sufficiently large $|c|$, this localization tendency holds for any value of $k$ for the lowest two bands. 
This observation leads us to expect that the Wannier functions of the first and second bands are localized at minima and maxima of Im[$V(x)$], respectively.
In fact, from each band, we can always construct the Wannier function which is localized and symmetric around a minimum or a maximum of Im[$V(x)$] by appropriately choosing the phases of the Bloch states and the unitary matrix $U(k)$, as discussed in Appendix.~\ref{sec:wannier}.
In Fig. \ref{fig:wannier-function} (a), we plot the Wannier functions for a large $|c|=200$, as described in Appendix.~\ref{sec:wannier}, that is, with $U_{11}(k)=e^{ik/4},\,U_{22}(k)=e^{-ik/4}, U_{12}(k)=U_{21}(k)=0$, and  $u_{l=0}^n (k)$, which are the $0$-th Fourier components of $u_k^n(x)$ defined above Eq. (\ref{eq:blochmatrix}), being real and positive. 
From Fig. \ref{fig:wannier-function} (a), we can understand that the obtained Wannier functions are indeed localized at the expected positions. 
This localization at a minimum and a maximum of the potential is the origin of the appearance of two lowest-energy bands. 
In the Hermitian case, we only obtain Bloch states localized at the minima of the scalar potential thus leading to the single lowest energy band.

\begin{figure}[tbp]
    \centering
    \includegraphics[width=6cm]{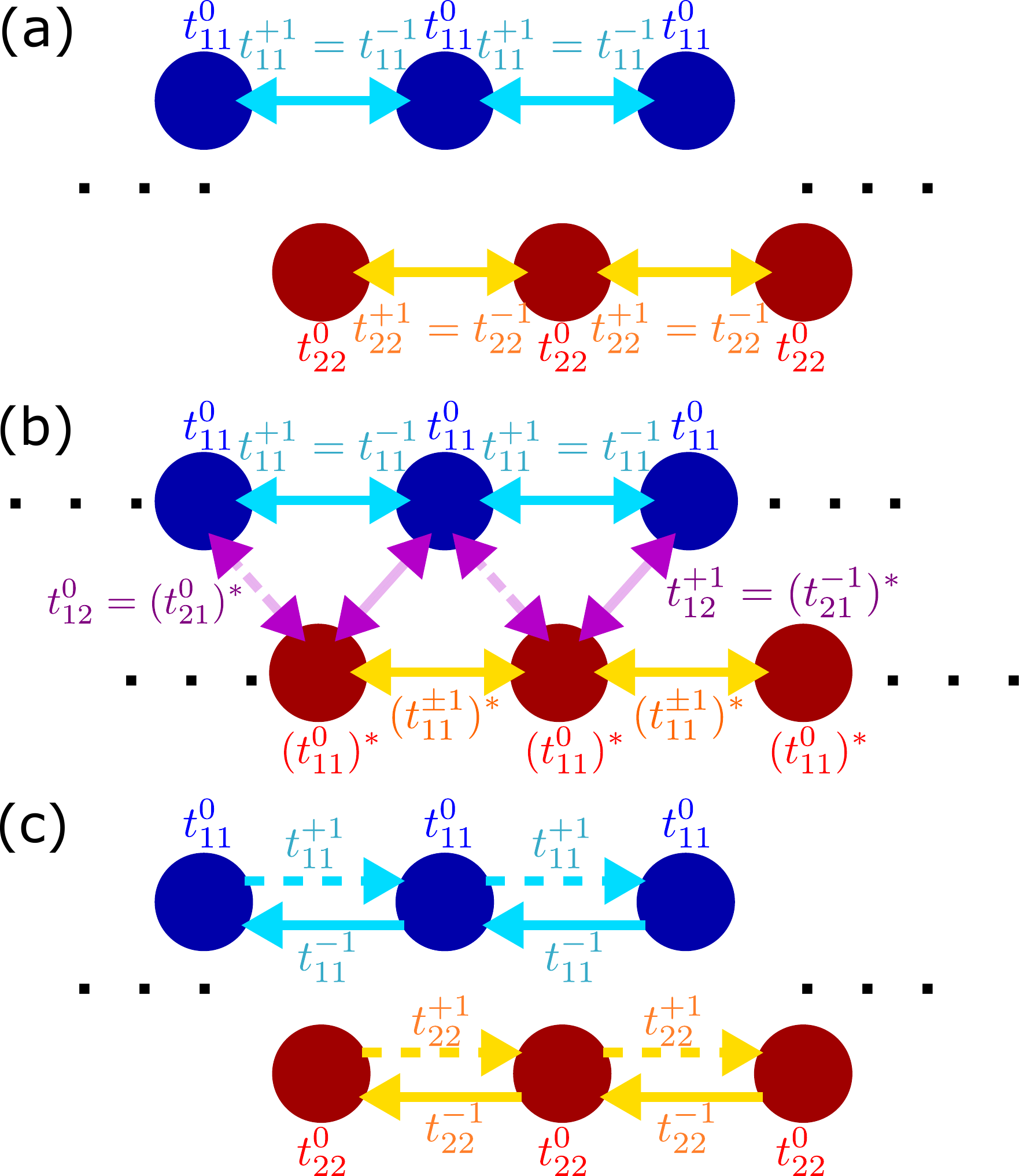}
    \caption{Schematic pictures which show the tight-binding lattices corresponding to our model for various regimes.
    (a) when two bands are separated and $A=0$,
    (b) when two bands are not separated and $A = 0$, and (c) when two bands are separated and $A\neq0$. 
    }
    \label{fig:lattice}
\end{figure}

Since the Wannier function of each band is constructed only from the Bloch states of each band, the tight-binding matrix elements for each band are given by
\begin{align}
    t_{nn}^{m-m^\prime} &= \frac{1}{N}\sum_{k}\varepsilon_n (k)e^{-ik (m-m^\prime)}
\label{eq:hopping_singleband}
\end{align}
with no inter-band terms, $t_{12}^{m-m^\prime} = 0$.
There are simple relations between the tight-binding matrix elements.
Since the energy eigenvalues of the lowest two bands obey the relation $\varepsilon_1 (k) = \varepsilon_2^\ast (k)$ as noted above and $\varepsilon_n (k) = \varepsilon_n (-k)$ as proven in Appendix.~\ref{sec:wannier}, the tight-binding matrix elements satisfy 
\begin{align}
    t_{11}^{m-m^\prime} = t_{11}^{m^\prime - m} = (t_{22}^{m-m^\prime})^* = (t_{22}^{m^\prime-m})^*.
    \label{eq:symmetry_singleband}
\end{align}
We note that the same relation holds also when two bands are mixed, but its proof is more involved as we discuss later. 
We consider the tight-binding approximation which includes hoppings only up to nearest neighbors. 
Then, making use of the above relation, the lowest two bands are described by two complex parameters
\begin{align}
    t \equiv t_{11}^{1} = t_{11}^{-1} = (t_{22}^{1})^* = (t_{22}^{-1})^*,\,    \gamma \equiv t_{11}^0 = (t_{22}^0)^*.
    \label{eq:parameters_truncated}
\end{align}
The resulting tight-binding lattice model is two decoupled chains, one for the first band and the other for the second band, as schematically depicted in Fig. \ref{fig:lattice} (a).
The corresponding dispersion relations with the tight-binding approximation are $\varepsilon_1 (k) = \gamma + 2t\cos (k)$ and $\varepsilon_2 (k) = \varepsilon_1^\ast(k)$, which agree well with the numerically obtained dispersion relations from the continuum model, as described in Fig. \ref{fig:eigenvalue_dispersion_pt} (e).

\begin{figure}[tbp]
    \centering
    \includegraphics[width=\columnwidth]{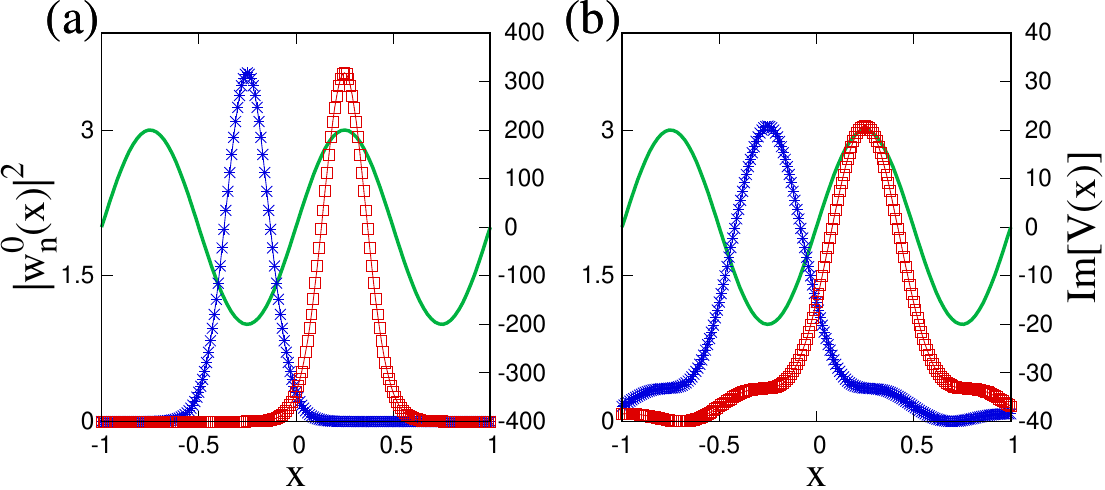}
    \caption{Wannier functions $w_1^0(x)$ (blue stars) and $w_2^0(x)$ (red squares) when $A = 0$ and $V(x) = c\sin (2\pi x)$ with (a) $c=200i$ and  (b) $c=20i$. 
    While the Wannier functions in (a) is obtained from individual band, (b) is based on mixing of the two bands as explained in the main text. 
    The green lines show the imaginary potentials $\mathrm{Im}[V(x)]$.}
    \label{fig:wannier-function}
\end{figure}

\subsubsection{Intermediate values of $|c|$: Wannier functions constructed by mixing two bands}
\label{subsubsec:intermediate-c}

We have seen that, in the limit of large $|c|$, the single-band tight-binding approximation nicely describes the lowest two bands. As the strength of the potential $|c|$ is weakened, we expect that there appear some hoppings between the Wannier functions localized at minima and maxima of Im[$V(x)$], namely there appear couplings between the two bands.
Below we show that this expectation is indeed correct, and the development of band structure of two lowest bands shown in Fig.~\ref{fig:eigenvalue_dispersion_pt}(c)-(e), such as the collision of exceptional points leading to the gap opening, can be well reproduced by the two-band tight-binding model.
As we show below, we find that the couplings between two bands are zero when the two bands are separated in the complex plane; the couplings between the two bands appear only when two bands are degenerate at some points in momentum space forming exceptional points.

We first discuss how we obtain localized Wannier functions when the two bands potentially mix, namely when the unitary matrix $U(k)$ can be a two-by-two matrix with finite off-diagonal terms $U_{12}(k)\neq0,\,U_{21}(k)\neq0$.
To obtain a good tight-binding description, we need to choose appropriate $U(k)$ so that Wannier functions are well localized at the expected positions.
To this end, we choose the construction of $U(k)$ based on the trial basis functions, utilizing the method which has been developed in the construction of multi-band Wannier functions in Hermitian Hamiltonians~\cite{marzani1997maximally,
souza2001maximally,
marzani2012maximally}.
Since we want the constructed Wannier functions to approach the ones obtained from individual bands in the large $|c|$ limit, such as the ones in Fig.~\ref{fig:wannier-function} (a), we choose trial bi-orthogonal functions $\{g_n(x)\}$ and $\{\tilde{g}_n(x)\}$ to be Wannier functions constructed from individual bands in the large $|c|$ case. We note that these trial functions are localized with their centers at a minimum or a maximum of Im[$V(x)$], where we want the constructed Wannier functions to be localized around.
Based on these trial basis functions, we choose the unitary matrix as
\begin{align}
    U(k)=D(k)[D^\dagger(k)D(k)]^{-\frac{1}{2}},\ 
    D_{nn'}(k)=\langle\tilde{\psi}_k^n|g_{n'}\rangle.
    \label{eq:guess-method}
\end{align}
Carrying out the singular value decomposition of $D(k)$, we see that $D(k)[D^\dagger(k)D(k)]^{-\frac{1}{2}}$ is a unitary matrix. 
We note that this construction also has an advantage that the resulting Wannier functions are independent of the phases of Bloch wavefunctions we choose.
Figure \ref{fig:wannier-function} (b) shows Wannier functions when $c=20i$ with this construction.
The constructed Wannier functions for $c = 20i$ are more spread than the Wannier functions for $c=200i$, but they are still centered around the minimum and maximum of Im[$V(x)$]. 
From these Wannier functions, we can construct the tight-binding model by truncating the long-range hoppings. 
We first note that the values of hoppings $t_{nn'}^{m-m'}$ are almost independent of the trial Wannier functions as long as the trial Wannier functions are calculated with sufficiently large $|c|$ and thus $\{g_n(x)\}$ are well localized; we find that the difference of $|t_{nn'}^{m-m'}|$ is only around 1\% when we choose $\{g_n(x)\}$ as Wannier functions with $c=200i$ and $400i$. 
When the Wannier functions are constructed with the linear combinations of the lowest two bands, there can be hoppings among Wannier functions localized at the minimum and the maximum of Im[$V(x)$]. 
The resulting tight-binding lattice model is a one-dimensional triangular ladder, as schematically depicted in Fig.~\ref{fig:lattice} (b), where blue and red sites correspond to Wannier states localized at the minima and maxima of Im[$V(x)$], respectively.

We can derive simple relations between hopping amplitudes utilizing symmetries present in the system. We make use of the $\mathcal{PT}$ symmetry of the Hamiltonian, $H_{-x}^\ast = H_{x}$, which implies $\varepsilon_1 (k)$ and $\varepsilon_2 (k)$ are either both real or complex conjugate pairs, and also $\varepsilon_n(k)=\varepsilon_n(-k)$. 
The $\mathcal{PT}$ symmetry also implies $g_1^* (-x) = g_2 (x)$ for the trial functions.
These properties are shown in Appendix.~\ref{sec:wannier}.
Using these properties, we can obtain
\begin{align}
    \begin{array}{cc}
    t^{m-m^\prime}_{11} = t^{m^\prime - m}_{11} = (t^{m-m'}_{22})^* =(t^{m'-m}_{22})^*,\\
    t^{m-m'}_{12} =(t^{m'-m}_{21})^\ast
    \end{array}
    \label{eq:relation_hoppings_pt}
\end{align}
where the detailed derivation is given in Appendix.~\ref{sec:hopping}. 
The relation $t^{m-m'}_{12} =(t^{m'-m}_{21})^\ast$ implies that the inter-band couplings are Hermitian, while the intra-band couplings are non-Hermitian in general. 
We note that these relations are satisfied for general $\mathcal{PT}$ symmetric systems, such as those where $V(x)$ includes the $\cos(2\pi x)$ term with a real coefficient. 
Furthermore, using the property that our trial wavefunctions are symmetric around $x=\pm1/2$, $g_n[-x+(-1)^n/2]=g_n(x)$ originating from $H_{-x\pm1/4}=H_x$, which are shown in Appendix.~\ref{sec:wannier}, we can obtain
\begin{align}
    t^{m-m'}_{12} = t_{12}^{m^\prime - m +1}
    \label{eq:relation_hoppings_pm1/4}
\end{align}
whose derivation is also given in Appendix.~\ref{sec:hopping}. 

Equation (\ref{eq:relation_hoppings_pt}) indicates that the Hamiltonian of the tight-binding model including the two lowest bands is
\begin{align}
    H_\text{t}(k)=\left[\begin{array}{cc}
        t_1(k) & t_2(k) \\
        t_2^\ast(k) & t_1^\ast(k)
    \end{array}\right]
    \label{eq:pt-symmetric_lattice-hamiltonian}
\end{align}
where $t_1(k)=\sum_mt^m_{11}e^{ikm}$ and $t_2(k)=\sum_mt^m_{12}e^{ikm}$ and the summation should be truncated according to the degree of approximation one wants.
From Eq. (\ref{eq:pt-symmetric_lattice-hamiltonian}), we can understand that the tight-binding model also satisfies $\mathcal{PT}$ symmetry $\sigma_xH_\text{t}^\ast(k)\sigma_x=H_\text{t}(k)$ where $\sigma_x$ is a Pauli matrix.

When the first and second bands are separated we can show $t_{12}^{m-m'} = 0$, whose proof is given in Appendix \ref{sec:hopping}, and thus the tight-binding model becomes two independent chains described in Fig. \ref{fig:lattice} (a).
The tight-binding model constructed from trial functions thus becomes equivalent to the tight-binding model constructed from individual bands when the two bands are separated, and reduces to the large $|c|$ case discussed above.
Figure \ref{fig:eigenvalue_dispersion_pt} (d) and (e) show the dispersion relations of the tight-binding model when the lowest two bands are separated and thus the corresponding tight-binding models are constructed from individual bands. 

When the lowest two bands are not separated and thus $t_{12}^{m-m'} \neq 0$, the tight-binding Hamiltonian in the momentum-space takes the following form
\begin{align}
    H_\text{t} (k)
    =
    \begin{pmatrix}
    \gamma + 2t\cos (k) & t^0_{12} (1 + e^{ik}) \\
    (t^0_{12})^* (1 + e^{-ik}) & \gamma^* + 2t^* \cos (k)
    \end{pmatrix},
        \label{eq:pt-symmetric_lattice-hamiltonian_truncated}
\end{align}
where only hoppings described in the triangular ladder in Fig.~\ref{fig:lattice} (b) are included.
The resulting dispersion relations for bands $n = 1,2$ become
\begin{align}
    \varepsilon_{n} (k) = &(-1)^n \sqrt{2|t_{12}^0|^2 [1 + \cos (k)] - (\mathrm{Im}[\gamma + 2t \cos (k)])^2} \notag \\
    &+\mathrm{Re}[\gamma + 2t\cos(k)].
\end{align}
Zeros of the first term determine the positions of the exceptional points in momentum space. 

\begin{figure}[tbp]
    \centering
    \includegraphics[width=8cm]{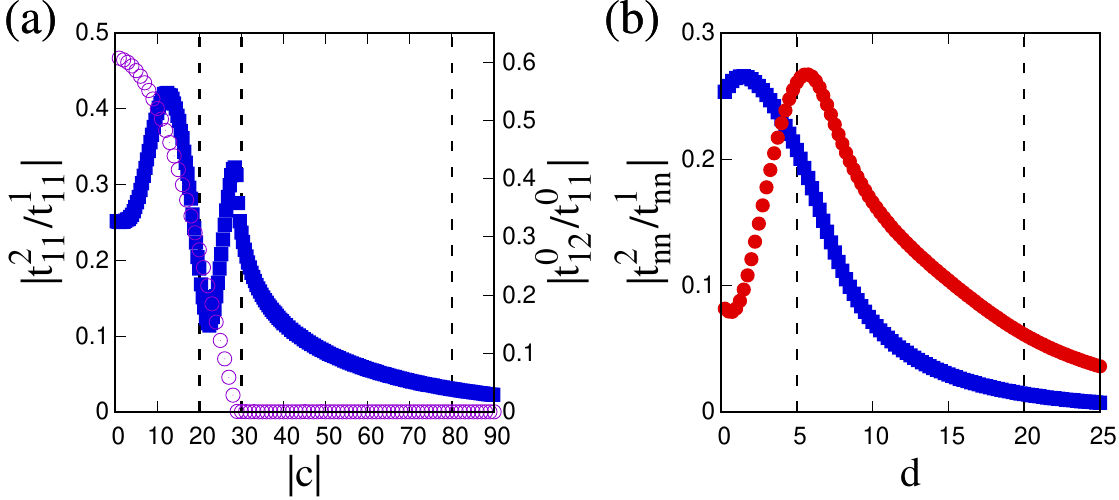}
    \caption{The ratio of hopping amplitudes as functions of the strength of the potential when $A=0$ and $V(x) = c\sin (2\pi x)$. 
    In (a), where $c$ is imaginary, blue filled squares and purple empty circles respectively correspond to $|t_{11}^2/t_{11}^1|$ and $|t_{12}^0/t_{11}^0|$. 
    Three dashed lines show parameters used in Fig. \ref{fig:eigenvalue_dispersion_pt} (c),\,(d), and (e). 
    In (b), where $c$ is complex and Re$(c)=d$ and Im$(c)=4d$, blue squares and red circles respectively show $|t_{11}^2/t_{11}^1|$ and $|t_{22}^2/t_{22}^1|$, as functions of $d$.  
    The left and right dashed lines correspond to parameters in Fig. \ref{fig:eigenvalue_dispersion_without-A} (a) and (b).}
    \label{fig:hopping-amplitude}
\end{figure}

Blue dashed lines in Fig. \ref{fig:eigenvalue_dispersion_pt} (c) show the dispersion relations under the tight-binding approximation based on the Wannier functions in Fig. \ref{fig:wannier-function} (b), from which we can understand that the dispersion relation of the continuum model is qualitatively well reproduced by the tight-binding model. 
Our two-band tight-binding model correctly accounts for the evolution of the dispersion relation as $|c|$ changes; when $|c|$ is small and thus $t_{12}^0 \neq 0$, the exceptional points appear, and as $|c|$ increases, the exceptional points collide and the two bands separate.
Further increasing $|c|$, the dispersion relations of the continuum model and the tight-binding model quantitatively agree, as shown in Fig. \ref{fig:eigenvalue_dispersion_pt} (e). 
The improved agreement of dispersion relations is due to the suppression of the long-range hopping terms. 
Blue filled squares in Fig. \ref{fig:hopping-amplitude} (a) show the ratio of the next-nearest to the nearest neighbor hopping amplitudes,  $|t_{11}^2/t_{11}^1|=|t_{22}^2/t_{22}^1|$, which becomes small as $|c|$ is increased. 
In the range $20\leq|c|\leq30$, $|t_{11}^2/t_{11}^1|$ exhibits a non-monotonic behavior, and it shows a peak around a gap-opening value of $|c|$.

\subsection{When $V(x) = c \sin (2\pi x)$ with complex $c$}
\label{subsec:complex-potential_without-A}
Next, we consider the situation where the coefficient $c$ of the scalar potential $V(x) = c \sin (2\pi x)$ has both real and imaginary parts. 
As we have seen, when $c$ is purely imaginary, the Wannier functions are localized at minima and maxima of $\sin (2\pi x)$. 
On the other hand, when $c$ is real and positive, the Wannier functions are localized only at the minima of $\sin (2\pi x)$. When $c$ has both real and imaginary parts, there is a competition between the real and imaginary parts. 
In Fig.~\ref{fig:bloch-function} (b) and (c), we plot the Bloch functions of the lowest two bands at $k = 0$ for different values of $\mathrm{Re}(c)$. 
We observe that, below a threshold value of $\mathrm{Re}(c)$, the Bloch states are localized both at minima and maxima of $\mathrm{Im}[V(x)]\propto\sin (2\pi x)$, similar to the case of purely imaginary $c$. 
However, above the threshold value, the Bloch states of the lowest two bands both become localized only at minima and one of the two Bloch states shows nodes at the minima. 
We can understand the localization of two Bloch states in minima of $\sin (2\pi x)$ from the limit of purely real $c$, where Wannier functions of the lowest two bands become $s$ and $p$ orbitals localized at the minima of $\sin (2\pi x)$; the nodal structure of one of the two Bloch states is in accordance with what we expect from the $p$-orbital Wannier function.

For constructing a tight-binding model to describe the lowest two bands, we note that, as mentioned in the nearly-free regime, exceptional points do not appear when $c$ has both real and imaginary parts. 
The lowest two bands are thus separated in the complex plane. 
We thus construct the tight-binding model from individual bands, without mixing the two bands. 
In this case, the tight-binding model becomes two independent chains as shown in Fig. \ref{fig:lattice} (a) and hopping terms are symmetric but non-Hermitian in general,
\begin{align}
    t_{nn}^{m-m'}=t_{nn}^{m'-m},
    \label{eq:hopping_symmetric}
\end{align}
which can be shown from $\varepsilon_n(-k)=\varepsilon_n(k)$ and Eq. (\ref{eq:hopping}) with diagonal $U(k)$. 
The dispersion relation from the tight-binding model, truncating the hopping up to the nearest neighbors, are plotted in comparison to the dispersion relation calculated from the continuum model in Fig.~\ref{fig:eigenvalue_dispersion_without-A}. 
The agreement improves as $|c|$ is increased. When $|c|$ is not large, as in Fig.~\ref{fig:eigenvalue_dispersion_without-A} (a), the influence of the higher bands is visible, showing the limitation of the tight-binding approximation in this regime.

We have also estimated the ratio of the next-nearest-neighbor to the nearest-neighbor hopping amplitudes as a function of the strength of the potential, fixing the ratio between the real and the imaginary parts $\mathrm{Im}(c)/\mathrm{Re}(c) = 4$.
The result is plotted in Fig. \ref{fig:hopping-amplitude} (b).
As expected, the next-nearest-neighbor hoppings become smaller as the strength of the potential is increased, which results in better agreement between the continuum model and the tight-binding approximation. We also notice that the next-nearest-neighbor hopping of the second band decays slower than that of the first band. 
We attribute this difference to the larger influence of higher bands to the second band. 

\subsection{When the vector potential is present}
\label{subsec:imaginary-potential_with-A}
When a vector potential $A$ is present, as presented in Fig.~\ref{fig:eigenvalue_dispersion_with-A}, the lowest two bands separate from the rest of the energy spectrum above a threshold value of $|c|$, for a purely imaginary scalar potential $V(x) = c\sin (2\pi x)$. Unlike the case when $A = 0$, the lowest two bands do not show mixing with exceptional points in momentum space. We therefore construct the tight-binding model without mixing the two bands, namely, we construct Wannier functions from individual bands. 
In the presence of the vector potential, the symmetric relation of Eq. (\ref{eq:hopping_symmetric}) does not hold any more because of $\varepsilon_n(-k)\neq\varepsilon_n(k)$. 
In particular, the hoppings become asymmetric $|t_{nn}^{m-m'}| \neq |t_{nn}^{m'-m}|$ just as in the Hatano-Nelson model as schematically described in Fig. \ref{fig:lattice} (c). Figure \ref{fig:eigenvalue_dispersion_with-A} (d) demonstrates that the tight-binding models obtained with this protocol well reproduce the dispersion relations of the continuum model when the strength of the scalar potential $|c|$ is large.

By adding a vector potential, we see the shift of the peaks of the Bloch wavefunction $\psi_k^n(x)$, as shown in Fig. \ref{fig:bloch-function} (d). 
In principle, the corresponding Wannier functions should be constructed to center around the shifted peaks by choosing phases of the Bloch states appropriately. 
However, as long as the tight-binding model is constructed from individual bands with diagonal $U(k)$, as in Eq. (\ref{eq:hopping_singleband}), the choice of the phases does not alter the hopping amplitudes and hence the resulting tight-binding model.

\section{Summary}
\label{sec:summary}
We have explored dispersion relations of continuum models under non-Hermitian periodic potentials. 

In the nearly-free regime where the strength of the imaginary scalar potential is small, we have found that the lowest two bands do not separate but form exceptional points.
This behavior is unique to imaginary periodic potentials because real periodic potentials open gaps with infinitesimally small strength, which is a well known fact in the ordinary Hermitian band theory \cite{ashcroft1976solid}.
In the presence of imaginary vector potentials, we found that the band separation is hindered when the scalar potential is small, and a different type of gap opens when the scalar potential is larger. 

When the scalar potential is strong, we can describe the band structures of the continuum model by discrete tight-binding models. 
The tight-binding models are constructed through not orthogonal but bi-orthogonal Wannier functions based on bi-orthogonal Bloch wavefunctions of the non-Hermitian Hamiltonian and its Hermitian conjugate. 

Traditionally, non-Hermitian physics has been developed in two opposite regimes; the nearly-free regime has been studied in relation to $\mathcal{PT}$-symmetric optics, whereas tight-binding models have been largely employed when topological structures of non-Hermitian models are discussed. 
Our work paves a way to connect these two regimes and provide a uniform understanding of non-Hermitian physics in a wide range of parameter spaces. 
With tight-binding models, various analyses become easier, such as the calculation of topological numbers and the derivation of $\mathcal{PT}$ symmetry breaking threshold. 
In this paper, we have focused on the simplest types of scalar and vector potentials, which serves as a first step toward understanding rich phenomena of the non-Hermitian band theory where tight-binding basis functions are constructed from the bi-orthogonal basis. 
Extending the work to more complicated periodic potentials, such as including internal degrees of freedom or considering scalar potentials with multiple minima/maxima per one period, to reach a more complete understanding of non-Hermitian band theory is left for future study. 
Extending the analysis to two or higher dimensional systems is also of great interest, in which case we need to consider a vector potential which is not just a constant, resulting in a complex magnetic field. 
In the Hermitian band theory, it is known that bands with nonzero Chern numbers do not give rise to localized Wannier functions~\cite{brouder2007}. We expect similar localization properties for Chern bands from non-Hermitian Hamiltonians, where we need to include multiple bands to construct localized bi-orthogonal Wannier basis.
Exploring the evolution of band structures and formation/collisions of exceptional points under these various types of non-Hermitian potentials, using both from the continuum theory and bi-orthogonal tight-binding models, will also shed light on further exploration of devices and phenomena inspired by non-Hermiticity in optics, acoustics, and other systems where the non-Hermitian Schr\"odinger equation can emerge.

\section{Acknowledgement}
This work was supported by JSPS KAKENHI Grant No. JP20H01845, JST PRESTO Grant No. JPMJPR19L2, JST CREST Grant No.JPMJCR19T1, and RIKEN iTHEMS.

\appendix

\section{Symmetric Wannier functions}
\label{sec:wannier}
We explain how we construct the Wannier functions which are symmetric around a minimum or a maximum of Im[$V(x)$], when the potential is $V(x) = c \sin( 2\pi x )$ with imaginary $c$. 

We construct the Wannier functions at the 0-th unit cell to be symmetric around $x = \pm1/4$, that is, we make Wannier functions fulfill $w_1^0(-x-1/2) = w_1^0 (x)$ and $w_2^0(-x+1/2) = w_2^0 (x)$. 
We demonstrate in detail how to construct the Wannier function $w_1^0(x)$ which is localized around $x = -1/4$ and symmetric around the localization center, which is one of the minima of Im[$V(x)]$. 
The construction of $w_2^0(x)$ symmetric around $x=+1/4$ can be carried out following the same procedure. 
What is crucial in the following argument is that the scalar potential has the same symmetry $V(-x-1/2) = V(x)$. 
We first show that $u_k^1(x)$ at opposite $k$ obeys
\begin{align}
    u_k^1(-x-1/2) = e^{i\theta_1(k)} u_{-k}^1(x), 
    \label{eq:uk_plus-minus}
\end{align}
that is, by flipping $u_k^1(x)$ around $x = -1/4$ one obtains $u_{-k}^1(x)$, apart from the overall phase factor when Bloch states are normalized. 
To show the relation in Eq.~(\ref{eq:uk_plus-minus}) we examine the structure of the eigenvalue equation in the matrix form, Eq.~(\ref{eq:blochmatrix}). 
Explicitly writing out $H_{lm}(k)$, this matrix equation in the absence of the vector potential is
\begin{align}
    \sum_m \left\{ (k+2\pi m)^2 \delta_{lm} + V_{l-m} \right\} u_m^n(k) = \varepsilon_n(k) u_l^n (k).
    \label{eq:appmatrix}
\end{align}
The symmetry of the scalar potential $V(-x-1/2) = V(x)$ implies
\begin{align}
    &V(x) = \sum_l V_l e^{i2\pi l x} = V(-x-1/2)\notag \\
    & = \sum_{l}V_l e^{-i \pi l} e^{-i2\pi l x} = \sum_{l}V_{-l}(-1)^{l} e^{i2\pi l x}.
\end{align}
Comparing the top and bottom lines, we obtain $V_l = V_{-l}(-1)^l$.
Using this relation to the equation obtained by flipping signs of $k$, $l$, and $m$ in Eq. (\ref{eq:appmatrix}), we obtain
\begin{align}
    \sum_m &\left\{ (k+2\pi m)^2 \delta_{lm} + V_{l-m}(-1)^{l-m} \right\} u_{-m}^n(-k) \notag \\
    &= \varepsilon_n(-k) u_{-l}^n (-k).
\end{align}
Multiplying both sides by $(-1)^l$, we obtain
\begin{align}
    \sum_m &\left\{ (k+2\pi m)^2 \delta_{lm} + V_{l-m} \right\} (-1)^{m} u_{-m}^n(-k) \notag \\
    &= \varepsilon_n(-k) (-1)^{l} u_{-l}^n (-k).
    \label{eq:potential_reflection}
\end{align}
Comparing this final equation with Eq.~(\ref{eq:appmatrix}), we see that the vectors $\{u_l^n(k)\}$ and $\{(-1)^{l} u_{-l}^n (-k)\}$ are the eigenvectors of the same matrix. 
Assuming that there is no degeneracy of energy, we can conclude that $\varepsilon_n (k) = \varepsilon_n (-k)$ and the eigenvectors $\{u_l^n(k)\}$ and $\{(-1)^{-l} u_{-l}^n (-k)\}$ are the same up to a phase factor provided that the eigenvectors are normalized
\begin{align}
    u_l^n(k) = e^{-i\theta_n(-k)}(-1)^{-l} u_{-l}^n (-k),
    \label{eq:fourier-components_reflected}
\end{align}
where $\theta_n(-k)$ is an $l$-independent phase factor. 
From this we can see the desired relation:
\begin{align}
    &u_k^n (-x-1/2) = \sum_l u_l^n(k)e^{-i2\pi l x}e^{-i\pi l}
    \notag \\
    &=
    \sum_l u_{-l}^n(k) (-1)^l e^{i2\pi l x}
    =
    \sum_l e^{i\theta_n(k)} u_l^n(-k) e^{i2\pi l x}
    \notag \\
    &=
    e^{i\theta_n(k)}u_{-k}^n (x).
    \label{eq:eq:uk_plus-minus_derivation}
\end{align}
In the same way, we can show $u_k^n(-x+1/2)=e^{i\theta_n(k)}u_{-k}^n(x)$. 
Using these transformation properties, we now show how we choose the phases of the Bloch states to construct the Wannier functions with the symmetries $w_1^0(-x-1/2) = w_1^0 (x)$ and $w_2^0(-x+1/2) = w_2^0 (x)$. 
From the definition of the Wannier function, we obtain
\begin{align}
    & w_1^0(-x-1/2)
    =
    \frac{1}{\sqrt{N}}\sum_{k,n} U_{n1}(k) \psi_k^n (-x-1/2) \notag \\
    &=
    \frac{1}{\sqrt{N}}\sum_{k,n} U_{n1}(k) e^{-ikx-ik/2}e^{i\theta_n(k)}u_{-k}^n(x)
    \notag \\
    &=
    \frac{1}{\sqrt{N}}\sum_{k,n} U_{n1}(-k) e^{ik/2} e^{i\theta_n(-k)} \psi_k^n (x).
    \label{eq:symmetric_w1}
\end{align}
Therefore, if we choose $U_{n1}(-k) e^{ik/2} e^{i\theta_n(-k)} = U_{n1}(k)$, the final line becomes equal to $w_1^0 (x)$ and the Wannier function respects the symmetry of the potential, $w_1^0(-x-1/2) = w_1^0 (x)$. 
In the same way, reflecting $w_2^0(x)$ around $x=+1/4$ results in
\begin{align}
    w_2^0(-x+1/2)=
    \frac{1}{\sqrt{N}}\sum_{k,n} U_{n2}(-k) e^{-ik/2} e^{i\theta_n(-k)} \psi_k^n (x),
    \label{eq:symmetric_w2}
\end{align}
and thus $w_2^0(-x+1/2)=w_2^0(x)$ is satisfied if matrix elements and phases are chosen as $U_{n2}(-k) e^{-ik/2} e^{i\theta_n(-k)} = U_{n2}(k)$. 
We note that there is a redundancy in defining the phase; one can include the phase $U_{nn}(k)$ in the definition of the Bloch state. 
Nevertheless, it is computationally useful to separate these two phases, one phase to be determined when we calculate the Bloch wavefunctions, and the other phase to be determined when constructing the Wannier function.
A choice of $e^{i\theta_n(k)}$ fixes the relative phase between Bloch states with opposite momenta $k$ and $-k$. 
There is still a freedom to choose relative phases of $u_k^n(x)$ with $k \ge 0$; as long as the relation $U_{nm}(-k) e^{-i(-1)^mk/2} e^{i\theta_n(-k)} = U_{nm}(k)$ is satisfied, choosing different phases for $u_k^n(x)$ with $k \ge 0$ yields different Wannier functions obeying the symmetry $w_n^0[-x+(-1)^n/2] = w_n^0 (x)$. 
One particular choice of the phase which we employed in numerically calculating the Bloch wavefunctions is to make $\theta_n(k) = 0$ and $u_{l=0}^n(k) \ge 0$. 
In the case that the Wannier functions are constructed from individual bands, or equivalently  $U_{12}(k)=U_{21}(k)=0$ is satisfied, we choose $U_{11}(k)=e^{+ik/4}$ and $U_{22}(k)=e^{-ik/4}$ which fulfills the conditions above and thus realizes symmetric Wannier functions around $x=\pm1/4$. 
We have also confirmed that this choice of phase yields Wannier functions well localized at $x = \pm1/4$ when $|c|$ is large, as shown in Fig. \ref{fig:wannier-function} (a). 
We expect that one can also apply the procedure of constructing maximally localized Wannier functions studied in Hermitian systems  \cite{marzani2012maximally,
marzani1997maximally,
souza2001maximally} to non-Hermitian systems. 
However, localization functions, which are to be minimized, can be defined either with respect to the right eigenstates or to the bi-orthogonal basis. 
Understanding the physical relevance of these two different localization functions and its consequence in the resulting tight-binding models are left for future works. 
When we construct $w_n^m(x)$ based on trial Wannier functions and $U(k)$ is not diagonal, $U_{n1}(-k)=e^{-ik/2}U_{n1}(k)$ and $U_{n2}(-k)=e^{+ik/2}U_{n2}(k)$ are also satisfied, as we clarify in Appendix \ref{sec:hopping}, which leads to symmetric Wannier functions around $x=\pm1/4$.

Finally we derive a relation between the Wannier functions localized around $x = -1/4$ and $x = +1/4$ which results from our phase convention and the $\mathcal{PT}$ symmetry of the Hamiltonian. 
The eigenvalue equation for the Bloch state of the first band takes the form
\begin{align}
    H_x e^{ikx}u_{k}^1(x) = \varepsilon_1 (k) e^{ikx} u_k^1 (x).
    \label{eq:eigen-equation_1}
\end{align}
Taking the complex conjugation of the above equation and making $x \to -x$, we obtain 
\begin{align}
    H_x e^{ikx} [u_{k}^1(-x)]^* = \varepsilon_1^* (k) e^{ikx} [u_k^1 (-x)]^*,
    \label{eq:eigen-equation_1_pt-pair}
\end{align}
where we used the $\mathcal{PT}$ symmetry of the Hamiltonian $H_{-x}^* = H_x$.
This relation shows that $\varepsilon_1^* (k)$ is also an eigenvalue of the Hamiltonian with momentum $k$. 
When $|c|$ is large and the two lowest bands are separated, the energy of the first band is not real and thus $\varepsilon_2 (k) = \varepsilon_1^* (k)$. 
Therefore, $e^{ikx}[u_k^1 (-x)]^*$, which is an eigenstate with the eigenvalue $\varepsilon_2 (k)$, should be the Bloch wavefunction of the second band with momentum $k$ up to a phase factor $e^{i\phi (k)}$. 
We then have an equality
\begin{align}
    u_k^2 (x) = e^{i\phi (k)} [u_k^1 (-x)]^*. \label{eq:uk2uk1}
\end{align}
Expanding both sides as $u_k^n (x) = \sum_l u_l^n(k) e^{i2\pi lx}$, the above relation implies
\begin{align}
    u_l^2 (k) = [u_l^1 (k)]^* e^{i\phi(k)}.
\end{align}
Since the phase factor $e^{i\phi(k)}$ is independent of $l$, we can fix the factor $e^{i\phi(k)}$ by examining this relation for $l = 0$. 
At $l=0$, our convention is to choose $u_{l=0}^2 (k)$ and $u_{l=0}^1 (k)$ to be real and positive, which implies $\phi(k) = 0$.
From $\phi(k) = 0$ and Eq. (\ref{eq:uk2uk1}), we can derive a useful relation for the Wannier functions.
The Wannier functions constructed from the first (second) band are localized at $x = -1/4$ ($x = +1/4$).
Then, our phase convention is to choose $U_{11}(k)= e^{ik/4}$ and $U_{22}(k)=e^{-ik/4}= U_{11}^*(k)$ when the first and second bands are separated. 
Then, we obtain
\begin{align}
    &[w_1^0 (-x)]^*
    =
    \frac{1}{\sqrt{N}}\sum_k U_{11}^\ast(k) e^{ikx} [u_k^1 (-x)]^*
    \notag \\
    &=
    \frac{1}{\sqrt{N}}\sum_k U_{22}(k) e^{ikx} u_k^2 (x)
    =
    w_2^0 (x).
\end{align}
This relation implies that by flipping the Wannier function constructed from the first band, which is centered around $x=-1/4$, and taking its complex conjugation, we obtain the Wannier function constructed from the second band, which is centered around $x = +1/4$. 
This relation will be useful in finding relations among tight-binding matrix elements, as discussed in the next section.

\section{Derivation of relations among tight-binding hopping amplitudes}
\label{sec:hopping}
Here, we derive various relations among tight-binding hopping amplitudes in Eq. (\ref{eq:relation_hoppings_pt}), by explicitly calculating $U(k)$ in Eq. (\ref{eq:guess-method}). 
When we write the singular value decomposition of $D(k)$ as
\begin{align}
    D(k)=E(k)F(k)G^\dagger(k)
    \label{eq:SVD_D}
\end{align}
where $E(k)$ and $G(k)$ are unitary matrices, $U(k)$ can be written as
\begin{align}
    U(k)=E(k)G^\dagger(k).
\end{align}
As mentioned in Appendix \ref{sec:wannier}, the trial Wannier functions under a strong imaginary potential are symmetric around their localization centers $x=\pm1/4$,
\begin{align}
    g_1(x)&=g_1(-x-1/2)=\frac{1}{\sqrt{N}}\sum_ke^{+i\frac{k}{4}}\psi_k^1(x),
    \label{eq:wannier-funcion_1}\\
    g_2(x)&=g_2(-x+1/2)=\frac{1}{\sqrt{N}}\sum_ke^{-i\frac{k}{4}}\psi_k^2(x). \label{eq:wannier-funcion_2}
\end{align}
Also, the Bloch functions form the $\mathcal{PT}$ symmetric pair $[\psi_k^1(-x)]^\ast=\psi_k^2(x)$ when the first and second bands are separated with imaginary $c$, resulting in the $\mathcal{PT}$ symmetric pair of trial Wannier functions
\begin{align}
    g_1^\ast(-x)=g_2(x).
    \label{eq:pt-pair_wannier-funcion}
\end{align}
For the calculation of $U(k)$, we separate the Brillouin zone into two regions $\alpha$ and $\beta$; $\varepsilon_n^\ast(k)=\varepsilon_n(k)$ in the region $\alpha$ and $\varepsilon_1^\ast(k)=\varepsilon_2(k)$ in the region $\beta$. 
In the region $\alpha$, if we choose phases of $\tilde{\psi}_k^n(x)$ such that $\tilde{u}_{l=0}^n(k)$ is real and positive, the Bloch eigenfunctions of $H_x^\dagger$ satisfy $\mathcal{PT}$ symmetry, 
\begin{align}
    [\tilde{\psi}_k^n(-x)]^\ast=\tilde{\psi}_k^n(x).
    \label{eq:pt-symmetric_alpha}
\end{align}
From Eqs. (\ref{eq:pt-pair_wannier-funcion}) and (\ref{eq:pt-symmetric_alpha}), we can understand that the matrix elements of $D(k)$ are related by
\begin{align}
    D_{11}(k)&=\int dx[\tilde{\psi}_k^1(x)]^\ast g_1(x)\nonumber\\
    &=\int dx\tilde{\psi}_k^1(-x)g_2^\ast(-x)=D_{12}^\ast(k)\equiv\alpha(k)
    \label{eq:relation_D_alpha_1}\\
    D_{21}(k)&=\int dx[\tilde{\psi}_k^2(x)]^\ast g_1(x)\nonumber\\
    &=\int dx\tilde{\psi}_k^2(-x)g_2^\ast(-x)=D_{22}^\ast(k)\equiv\alpha'(k)    \label{eq:relation_D_alpha_2}
\end{align}
and $D(k)$ can be written as
\begin{align}
    D(k)=\begin{pmatrix}
    \alpha(k)&\alpha^\ast(k)\\
    \alpha'(k)&[\alpha'(k)]^\ast
    \end{pmatrix}.
    \label{eq:D_alpha}
\end{align}
Through the singular value decomposition of $D(k)$ in Eq. (\ref{eq:D_alpha}), we can find that the components of  $U(k)=D(k)[D^\dagger(k)D(k)]^{-\frac{1}{2}}$ are
\begin{align}
    U_{11}(k)=&\frac{1}{2}\alpha^\ast(k)e^{-i\varphi_k}
    \left[\frac{1}{\sqrt{A_+(k)}}-\frac{1}{\sqrt{A_-(k)}}\right]\nonumber\\
    &+\frac{1}{2}\alpha(k)\left[\frac{1}{\sqrt{A_+(k)}}+\frac{1}{\sqrt{A_-(k)}}\right]
    \label{eq:U11_alpha}\\
    U_{21}(k)=&\frac{1}{2}[\alpha'(k)]^\ast e^{-i\varphi_k}
    \left[\frac{1}{\sqrt{A_+(k)}}-\frac{1}{\sqrt{A_-(k)}}\right]\nonumber\\
    &+\frac{1}{2}\alpha'(k)
    \left[\frac{1}{\sqrt{A_+(k)}}+\frac{1}{\sqrt{A_-(k)}}\right],
    \label{eq:U21_alpha}
\end{align}
where $U_{12}(k)=U_{11}^\ast(k),\,U_{22}(k)=U_{21}^\ast(k),\,e^{i\varphi_k}=([\alpha(k)]^2+[\alpha'(k)]^2)^\ast/|[\alpha(k)]^2+[\alpha'(k)]^2|$, and $A_\pm(k)=|\alpha(k)|^2+|\alpha'(k)|^2\pm|[\alpha(k)]^2+[\alpha'(k)]^2|$. 
Equations (\ref{eq:U11_alpha}) and (\ref{eq:U21_alpha}) indicate that all components of $U(k)$ have the same absolute value
\begin{align}
    |U_{ij}(k)|=\frac{1}{\sqrt{2}},\,\,\{i,j\}=\{1,2\}
    \label{eq:magnitude_U_alpha}
\end{align}
in the region $\alpha$. 
In the region $\beta$, $\mathcal{PT}$ symmetry of Bloch eigenfunctions is broken and thus
\begin{align}
    [\tilde{\psi}_k^1(-x)]^\ast=\tilde{\psi}_k^2(x)
    \label{eq:pt-broken_beta}
\end{align}
is satisfied with $\tilde{u}_{l=0}^n(k)$ being real and positive. 
From Eq. (\ref{eq:pt-broken_beta}), we can find 
\begin{align}
    D_{11}(k)&=\int dx[\tilde{\psi}_k^1(x)]^\ast g_1(x)\nonumber\\
    &=\int dx\tilde{\psi}_k^2(-x)g_2^\ast(-x)=D_{22}^\ast(k)\equiv\beta(k)
    \label{eq:relation_D_beta_1}\\
    D_{12}(k)&=\int dx[\tilde{\psi}_k^1(x)]^\ast g_2(x)\nonumber\\
    &=\int dx\tilde{\psi}_k^2(-x)g_1^\ast(-x)=D_{21}^\ast(k)\equiv\beta'(k) \label{eq:relation_D_beta_2}
\end{align}
and thus $D(k)$ can be written as
\begin{align}
    D(k)=\begin{pmatrix}
    \beta(k)&\beta'(k)\\
    [\beta'(k)]^\ast&\beta^\ast(k)
    \end{pmatrix}.
    \label{eq:D_beta}
\end{align}
Carrying out the singular value decomposition of $D(k)$ in Eq. (\ref{eq:D_beta}), we can find
\begin{align}
    U_{11}(k)=&\frac{1}{2}\beta'(k)e^{-i\varphi_k}
    \left[\frac{1}{\sqrt{B_+(k)}}-\frac{1}{\sqrt{B_-(k)}}\right]\nonumber\\
    &+\frac{1}{2}\beta(k)\left[\frac{1}{\sqrt{B_+(k)}}+\frac{1}{\sqrt{B_-(k)}}\right]
    \label{eq:U11_beta}
\end{align}
and
\begin{align}
    U_{12}(k)&=U_{21}(k)=0
    \label{eq:U12_beta}
\end{align}
are satisfied in the region $\beta$, with  $U_{22}(k)=U_{11}^\ast(k),\,e^{i\varphi_k}=\beta^\ast(k)\beta'(k)/|\beta(k)\beta'(k)|$, and $B_\pm(k)=|\beta(k)|^2+|\beta'(k)|^2\pm2|\beta(k)\beta'(k)|$.

We can derive the relations of $t_{nn'}^{m-m'}$ in Eq. (\ref{eq:relation_hoppings_pt})  from $U(k)$ in regions $\alpha$ and $\beta$ obtained through $\mathcal{PT}$ symmetry $H_{-x}^\ast=H_x$. 
Equations (\ref{eq:magnitude_U_alpha}) and (\ref{eq:U12_beta}) indicate that hopping amplitudes in Eq. (\ref{eq:hopping}) can be written as
\begin{align}
    &t_{nn}^{m-m'}=\sum_{k\in\alpha}\varepsilon_+(k)
    \frac{e^{-ik(m-m')}}{2N}+\sum_{k\in\beta}\varepsilon_n(k)\frac{e^{-ik(m-m')}}{N}
    \label{eq:hopping_diagonal}\\
    &t_{12}^{m-m'}=\sum_{k\in\alpha}U_{11}^\ast(k)U_{12}(k)
    \varepsilon_-(k)\frac{e^{-ik(m-m')}}{N}
    \label{eq:hopping_12}\\
    &t_{21}^{m-m'}=-\sum_{k\in\alpha}U_{22}^\ast(k)U_{21}(k)
    \varepsilon_-(k)\frac{e^{-ik(m-m')}}{N}
    \label{eq:hopping_21}
\end{align}
where $\varepsilon_+(k)=\varepsilon_1(k)+\varepsilon_2(k)$ and $\varepsilon_-(k)=\varepsilon_1(k)-\varepsilon_2(k)$. 
From Eq. (\ref{eq:hopping_diagonal}), we can understand that $t_{nn}^{m-m'}=t_{nn}^{m'-m}$ and $t_{11}^{m-m'}=(t_{22}^{m-m'})^\ast$ are satisfied owing to $\varepsilon_1^\ast(k)=\varepsilon_2(k)$ in the region $\beta$ and $\varepsilon_n(k)=\varepsilon_n(-k)$ in the whole Brillouin zone. 
Also, in the light of $U^\dagger_{11}(k)U_{12}(k)+U^\dagger_{12}(k)U_{22}(k)=0$, Eqs. (\ref{eq:hopping_12}) and (\ref{eq:hopping_21}) indicate $(t_{12}^{m-m'})^\ast=t_{21}^{m'-m}$. 
In addition, from Eqs. (\ref{eq:hopping_12}) and (\ref{eq:hopping_21}), we can find 
$t_{12}^{m-m'}=t_{21}^{m'-m}=0$ when the first and second bands are separated or equivalently the region $\alpha$ is absent. 
In this case, the tight-binding models based on trial Wannier functions become the same as the tight-binding models for individual bands, respectively discussed in Secs. \ref{subsubsec:intermediate-c} and \ref{subsubsec:large-c}.

From the matrix components of $U(k)$ in the regions $\alpha$ and $\beta$, we can also show that $t_{12}^{-m}=t_{12}^{m+1}$ in Eq. (\ref{eq:relation_hoppings_pm1/4}) is satisfied, through the symmetry around $x=\pm1/4$, $H_{-x\pm1/2}=H_x$. 
To this end, we first derive a relation between $D_{ij}(k)$ and $D_{ij}(-k)$. 
Since the trial Wannier function of the second band satisfies $g_2(x)=g_2(-x+1/2)$, $D_{n2}(-k)$ becomes
\begin{align}
    D_{n2}(-k)&=\int dxe^{ikx}[\tilde{u}_{-k}^n(x)]^\ast g_2(x)\nonumber\\
    &=\int dxe^{ik(-x+1/2)}[\tilde{u}_{-k}^n(-x+1/2)]^\ast g_2(-x+1/2)\nonumber\\
    &=e^{ik/2}\int dxe^{-ikx}[\tilde{u}_{k}^n(x)]^\ast g_2(x)\nonumber\\
    &=e^{ik/2}D_{n2}(k),
    \label{eq:reflection_D}
\end{align}
where we changed the variable of the integral from $x$ into $-x+1/2$ in the second line and used $\tilde{u}_{-k}(-x+1/2)=\tilde{u}_k(x)$ which was shown in Appendix \ref{sec:wannier}. 
In the same way, we can show $D_{n1}(-k)=e^{-ik/2}D_{n1}(k)$. 
With these relations for the components of $D(k)$, Eqs. (\ref{eq:U11_alpha}), (\ref{eq:U21_alpha}), and (\ref{eq:U11_beta}) indicate $U_{n1}(-k)=e^{-ik/2}U_{n1}(k)$ and $U_{n2}(-k)=e^{ik/2}U_{n2}(k)$. 
These relations for the components of $U(k)$ result in symmetric Wannier functions around $x=\pm1/4$ 
\begin{align}
    w_1^0(-x-1/2)=w_1^0(x),\,w_2^0(-x+1/2)=w_2^0(x)
    \label{eq:symmetric-wannier-functions}
\end{align}
even when first and second bands are not separated, which can be understood from Eqs. (\ref{eq:symmetric_w1}) and (\ref{eq:symmetric_w2}) with $\theta_n(k)=0$. 
The corresponding bi-orthogonal Wannier functions $\tilde{w}_1^0(x)$ and $\tilde{w}_2^0(x)$ satisfy the same symmetries of $w_1^0(x)$ and $w_2^0(x)$. 
Therefore, the hopping amplitudes between different bands satisfy 
\begin{align}
    t_{12}^{-m}&=\int dx\tilde{w}_1^0(x)H_xw_2^m(x)\nonumber\\
    &=\int dx\tilde{w}_1^0(-x-1/2)H_{-x-1/2}w_2^m(-x-1/2)\nonumber\\
    &=\int dx\tilde{w}_1^0(x)H_xw_2^{-m-1}(x)=t_{12}^{m+1}
    \label{eq:hopping_symmetric_pm1/4}
\end{align}
which corresponds to Eq. (\ref{eq:relation_hoppings_pm1/4}), since the reflection around $x=-1/4$ corresponds to the reflection around $x=-m-1+1/4$ after the translation of $-2m-1$ for $w_2^m(x)$.

\bibliography{reference.bib}
\end{document}